\documentclass[aps,prx,reprint,nofootinbib,superscriptaddress]{revtex4-2}

\usepackage[utf8]{inputenc} 
\usepackage[T1]{fontenc}
\usepackage{graphicx, comment}
\usepackage{amsfonts}
\usepackage{amsmath,amssymb,bbm}
\usepackage{bm}
\usepackage{color}
\usepackage{epstopdf}
\usepackage{makecell}
\usepackage{subcaption}
\usepackage{epsfig}
\usepackage{verbatim}
\usepackage{lineno}
\usepackage[thicklines]{cancel}
\usepackage{url}   
\usepackage{xcolor}
\usepackage{listings}
\usepackage{multirow}
\usepackage{float}
\makeatletter
\let\newfloat\newfloat@ltx
\makeatother
\usepackage{algorithm,algpseudocode}
\usepackage{enumerate}
\usepackage{siunitx}
\usepackage{booktabs}
\usepackage{hyperref}
\hypersetup{
     colorlinks   = true,
     citecolor    = blue
}

\graphicspath{ {./} }
\usepackage{textcomp}
\usepackage{color}

\definecolor{red}{rgb}{1,0.,0}

\newcommand{\belem}{\texttt{ibmq\_belem}}
\newcommand{\jakarta}{\texttt{ibmq\_jakarta}}
\newcommand{\guadalupe}{\texttt{ibmq\_guadalupe}}
\newcommand{\toronto}{\texttt{ibmq\_toronto}}

\newcommand{\scriptG}[1]{\ensuremath{\mathcal{#1}}}
\newcommand{\G}[2]{\ensuremath{\mathcal{#1}_\mathrm{{#2}}}}
\newcommand{\Gstate}[3]{\ensuremath{|\psi_{\mathcal{#1}}\rangle = |{#2},{#3}\rangle}}

\newcommand{\naive}[0]{\textit{Na\"{i}ve}}
\newcommand{\unitary}[0]{\textit{Unitary}}

\begin{document}
\title{An entanglement-based volumetric benchmark for near-term quantum hardware}
\thanks{This manuscript has been authored by UT-Battelle, LLC under Contract No. DE-AC05-00OR22725 with the U.S. Department of Energy. The United States Government retains and the publisher, by accepting the article for publication, acknowledges that the United States Government retains a non-exclusive, paid-up, irrevocable, world-wide license to publish or reproduce the published form of this manuscript, or allow others to do so, for United States Government purposes. The Department of Energy will provide public access to these results of federally sponsored research in accordance with the DOE Public Access Plan
(http://energy.gov/downloads/doe-public-access-plan).}

\author{Kathleen E.\ Hamilton}
\email{hamiltonke@ornl.gov}
\affiliation{Quantum Computational Science Group,\ Oak\ Ridge\ National\ Laboratory,\
  Oak\ Ridge,\ TN,\ 37831,\ USA}
\affiliation{Computer Science and Engineering Division,\ Oak\ Ridge\ National\ Laboratory,\ Oak\ Ridge,\ TN,\ 37831,\ USA}
\author{Nouamane Laanait}
\affiliation{Carelon Digital Platforms, \
740 West Peachtree Street NW,\
Atlanta, GA 30308}

\author{Akhil Francis}
\affiliation{Department of Physics,\
  North Carolina State University,\
  Raleigh, NC 27695 USA}

\author{Sophia E.\ Economou}
\affiliation{Department of Physics,\
  Virginia Tech,\
  Blacksburg, VA USA}

\author{George S.\ Barron}
\affiliation{Department of Physics,\
  Virginia Tech,\
  Blacksburg, VA USA}
  
\author{K\"ubra Yeter-Aydeniz}
\affiliation{Emerging Engineering and Physical Sciences Department,\
  The MITRE Corporation,\
  McLean, VA, 22102-7539 USA}

\author{Titus Morris}
\affiliation{Quantum Computational Science Group,\ Oak\ Ridge\ National\ Laboratory,\
  Oak\ Ridge,\ TN,\ 37831,\ USA}
\affiliation{Computer Science and Engineering Division,\ Oak\ Ridge\ National\ Laboratory,\ Oak\ Ridge,\ TN,\ 37831,\ USA}

\author{Harrison Cooley}
\affiliation{Department of Physics,\
  Georgetown University,\
  Washington, DC, 20007 USA}

\author{Muhun Kang}
\affiliation{Department of Physics,\
  Cornell University,\
  Ithaca, NY, 14583 USA}
  
\author{Alexander F. Kemper}
\affiliation{Department of Physics,\
  North Carolina State University,\
  Raleigh, NC 27695 USA}

\author{Raphael Pooser}
\affiliation{Quantum Computing and Sensing Group,\ Oak\ Ridge\ National\ Laboratory,\
  Oak\ Ridge,\ TN,\ 37831,\ USA}
\affiliation{Computer Science and Engineering Division,\ Oak\ Ridge\ National\ Laboratory,\ Oak\ Ridge,\ TN,\ 37831,\ USA}

\begin{abstract}
We introduce a volumetric benchmark for near-term quantum platforms based on the generation and verification of genuine entanglement across n-qubits using graph states and direct stabilizer measurements. Our benchmark evaluates the robustness of multipartite and bipartite n-qubit entanglement with respect to many sources of hardware noise: qubit decoherence, CNOT and swap gate noise, and readout error. We demonstrate our benchmark on multiple superconducting qubit platforms available from IBM (\belem{}, \toronto{}, \guadalupe{} and \jakarta{}). Subsets of $n<10$ qubits are used for graph state preparation and stabilizer measurement.  Evaluation of genuine and biseparable entanglement witnesses we report observations of $5$ qubit genuine entanglement, but robust multipartite entanglement is difficult to generate for $n>4$ qubits and identify two-qubit gate noise as strongly correlated with the quality of genuine multipartite entanglement.
\end{abstract}

\maketitle
\section{Introduction}
\label{sec:introduction}

Developing benchmarks for quantum processors is crucial to advance the state of the art in noisy intermediate-scale quantum (NISQ) computing. Meaningful hardware benchmarks need to be applicable to different qubit technologies, and adapt to devices with different qubit layouts and basis gate sets.  Application and algorithm metrics need to quantify the complexity of the states that can be generated in large Hilbert spaces, and assist in developing intuition behind the effect of hardware noise and performance. 
While it is difficult to distill the complex operation of a quantum processor to a single number or a small set of numbers, a good benchmark should accurately capture the relative performance of different processors; it should be as simple as possible, and it should have reasonable scaling as a function of the system size.
A comparison of the proposed benchmarks found in the current literature shows the trade-off between scaling and detailed state characterization.
Tomographic methods \cite{kimmel2014_RB,waegell_benchmarks_2019} obtain highly detailed representation of quantum states and processes, but scale exponentially. Low-level benchmarks characterize hardware elements (e.g., qubits and gates) but interpolating from low-level to high-level (application) performance may not be straightforward. On the other hand, high-level benchmarks may fail to capture the salient features of the processors. 

A unifying framework for hardware characterization is provided by the Volumetric Benchmarks (VBs). IBM introduced quantum volume \cite{cross2019validating}, which benchmarks quantum hardware using randomized circuits. Hardware is assigned a scalar value based on the largest circuit that can be successfully executed, where success is defined by the measurement of heavy output from a random circuit.
The randomized circuits used to evaluate quantum volume are characterized as \textit{square}: where the width (number of qubits) and depth (number of computational steps) are equal. This framework was generalized in Ref. \cite{blume2020volumetric} to rectangular random circuits (where circuit depth can exceed width). Quantum volume as originally proposed is not clearly related to algorithmic performance or structured entanglement measures, although further refinements have become more closely related to algorithm performance \cite{blume2020volumetric}.

Quantum entanglement has been shown to be a valuable resource in quantum sensing and metrology \cite{giovannetti2011advances}, quantum communications \cite{pirker2018modular,epping2017multi} and quantum error correction \cite{rodriguez-blanco_efficient_2021}
when computing with pure quantum states \cite{jozsa2003role}. 
It is therefore reasonable to design benchmarks based on the ability of a processor to generate entangled states. For example, an interesting--but potentially oversimplified --- benchmark is the ability to produce Greenberger–Horne–Zeilinger (GHZ) states \cite{guhne_toolbox_2007,mainiero_homological_2019,wei_verifying_2020,mooney_generation_2021}. While GHZ states have been used as a low-level coherent noise characterization tool~\cite{zhou_entanglement_2020}, they only represent one specific class of entanglement out of the plethora of entangled states of $n$ qubits. 
Beyond GHZ states, specific cluster states and arbitrary graph states have been proposed as benchmarks of near-term quantum hardware ~\cite{nutz_efficient_2017,zhou_detecting_2019}.  
While a number of studies have used GHZ, cluster state or graph state preparation as a standalone benchmark \cite{wang_16qubit_2018,mooney_entanglement_2019,mooney_whole-device_2021}, individual state preparation as a benchmark does not generally give a global enough picture of the device's capabilities in quantum information processing. Yet, these observations of multipartite entanglement on contemporary processors do demonstrate that NISQ devices are capable of storing large entangled states, further motivating the need for a standard entanglement-based benchmark~\cite{varela_enhancing_2022,eldredge_entanglement_2020,chen_how_2022,dupont_entanglement_2022}.

The challenge in generalizing and improving entanglement-based benchmarks is that there exist too many entangled states of $n$ qubits. More importantly,
the role of entanglement as a computational resource is subtle, especially when generic entangled states are considered \cite{gross2009most,bremner2009random}. 
As a result, preparing a small number of arbitrary entangled states does not allow for a meaningful 
benchmark of hardware capabilities, and may have minimal use in both near-term and long-term quantum computing.

\begin{figure}[htbp]
    \centering
    \includegraphics[width=\columnwidth]{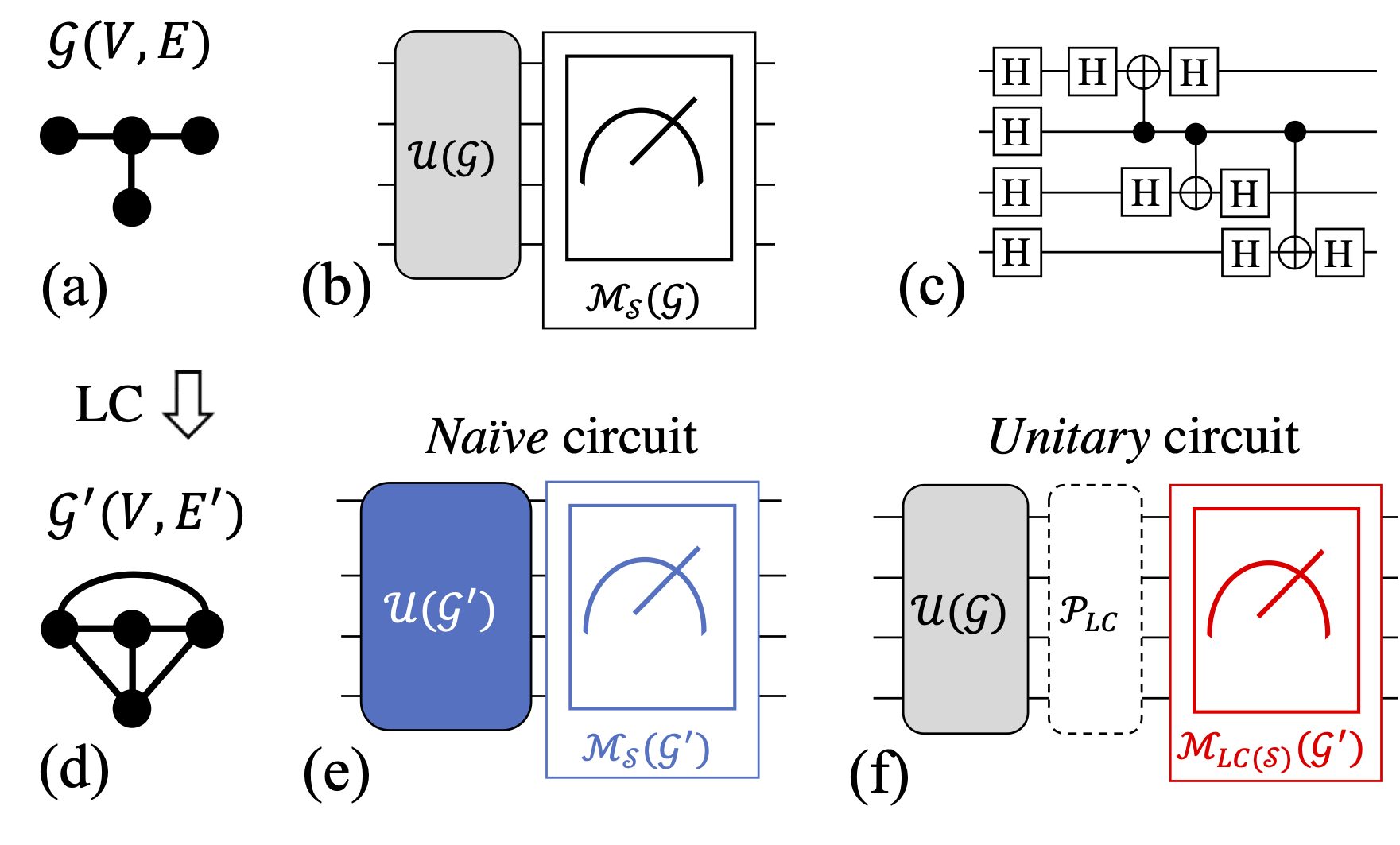}
    \caption{\textbf{Overview of the robust entanglement score} (a) the graph \G{G}{}{(V,E)} which defines (b) the unitary $\mathcal{U}(\mathcal{G})$ that prepares the graph state \Gstate{G}{n}{E} and the stabilizers $\mathcal{M}_{\mathcal{S}}(\mathcal{G})$ for the entanglement witness. (c) The unitary is constructed using Hadamard gates and CNOT gates. (d) Applying a local complement to \G{G}{}{(V,E)} defines $\mathcal{G}^{\prime}(V,E^{\prime})$. (e,f) The associated graph state can be prepared using
    (e) the \textit{\naive{}} method which constructs a new graph unitary $\mathcal{U}(\mathcal{G^{\prime}})$ using the edge set $E^{\prime}(\mathcal{G}^{\prime})$ or (f) the \textit{\unitary{}} method which appends single qubit operations $\mathcal{P}_{LC}$ to implement the LC operation on the original graph unitary $\mathcal{U}(\mathcal{G})$.}
    \label{fig:graph_state_circuit_construction}
\end{figure}
In this work we introduce a VB for entangled state generation, \textit{robust entanglement score} (RES), that encompasses practically relevant classes of entangled states and also scales linearly with circuit width. Its intended goal is to show robust genuine entanglement \cite{vidal1999robustness} over a large random sampling of entangled quantum states that
are used for a variety of quantum information tasks, including quantum error correction, sensing, and one-way computing. At the heart of our new benchmark is the generation and verification of $n$-qubit entangled \textit{graph} states that are subgraphs of the device connectivity map, and their locally equivalent graph states (Fig.~\ref{fig:graph_state_circuit_construction}(a),(d)). 
By choosing graph states that are amenable to the device connectivity map, the state preparation circuits can be constructed with minimal two-qubit gates (Fig.~\ref{fig:graph_state_circuit_construction}(c)). But, by generating graph states in the same equivalence class, we test graph states that are not guaranteed to have low gate overhead, or that may require measurement in different bases. 
Graph states in the same equivalence class can have very different graph structure, from sparse to dense (e.g. containing high degree vertices, triangles). 

Entanglement verification is the measure of success for our VB, but successful verification relies on high-fidelity stabilizer measurements that reveal the underlying hardware quality.  Thus, through our benchmark we characterize hardware performance that can be used to infer algorithmic performance where the underlying circuit structure is graph-like or for applications that make extensive use of entanglement (e.g. quantum optimization or physics simulations on lattices). 
Our results show that a benchmark such as RES, which exposes both global and subtle information, and allows for the direct evaluation of application-driven performance, is necessary in capturing and comparing device performance compared to other metrics. 

Overall, our benchmark evaluates hardware performance using graph state preparation, independent stabilizer measurements, and quantification of multi-qubit entanglement through the evaluation of genuine and biseparable witnesses \cite{toth_detecting_2005,jungnitsch_entanglement_2011}. The benchmark leverages specific graph state properties: local complementation of graphs and efficient state characterization via entanglement witnesses \cite{bourennane_multipartite_2004,jungnitsch_entanglement_2011,Jungnitsch_multiparticle_2011}.  This results in global information about the processor performance and its ability to sustain genuine multipartite entanglement among qubit subsets of the processor.  These aspects are presented in Section \ref{sec:methods}. We execute our benchmark on superconducting (SC) qubits, but in general our benchmark is hardware-agnostic and can be used to compare different types of devices.  Our benchmark is also application agnostic, though graph states are commonly utilized in many
quantum applications and act as an algorithmic primitive (see Section \ref{sec:discussion}).

\begin{figure*}[htbp]
    \centering
    \includegraphics[width=\textwidth]{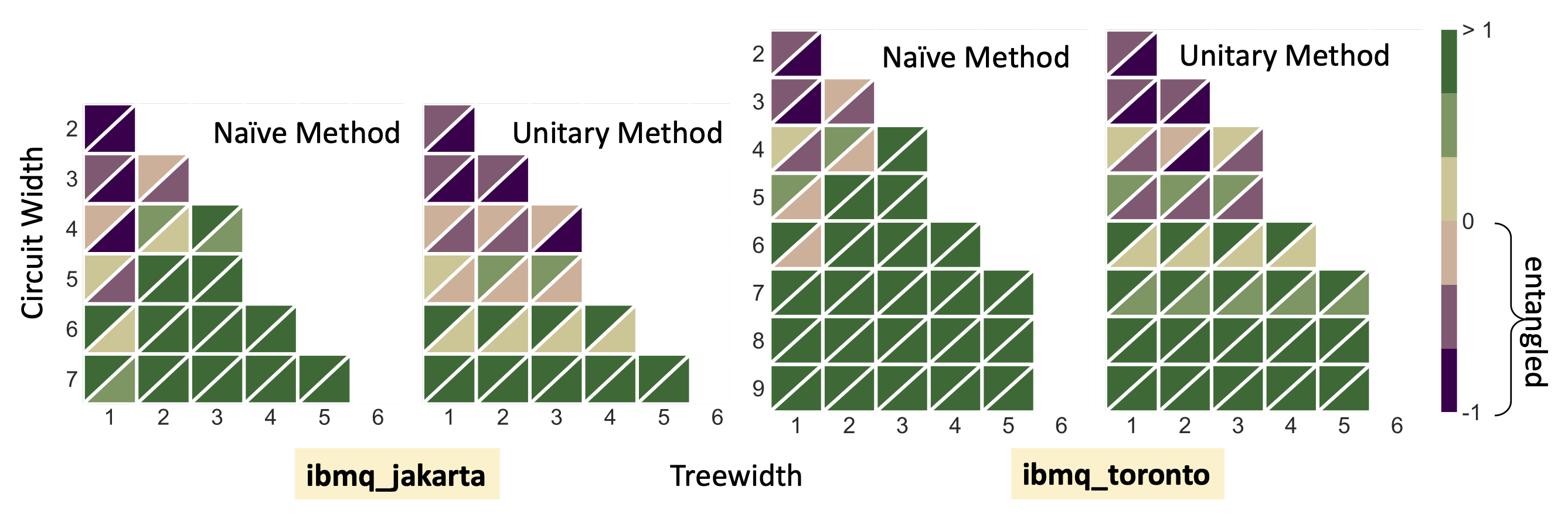}
    \caption{Median genuine witness values aggregated over multiple graph state executions. The upper triangular values are defined by non-mitigated results, the lower triangular values are defined by error mitigated results, using a tensored mitigator defined in Qiskit \texttt{Ignis}. Negative values indicate that entanglement.(Left) Benchmarking of \jakarta{} (RES-N$=6$, RES-U$=12$).  (Right) Benchmarking of \toronto{} (RES-N$=6$, RES-U$=8$).}
    \label{fig:jakarta_toronto_genuine_heatmaps}
\end{figure*}
\subsection*{Graph States and Entanglement Witnesses}
\label{sec:witness_methods}
The edges $E$ and vertices $V$ of a simple graph \G{G}{}{(V,E)} define a unitary that prepares the $n = |V|$ qubit graph state $\Gstate{G}{n}{E} = \mathcal{U}(\mathcal{G})|0\rangle^{\otimes n}$ using two-qubit entangling operations specified by the edge set $E(\mathcal{G})$. We verify the generation of multipartite and bipartite entanglement using witness functions ($\mathcal{W}$) constructed from the graph state stabilizers \G{M}{S}{($\mathcal{G}$)} \cite{toth_detecting_2005,toth_entanglement_2005,jungnitsch_entanglement_2011,Jungnitsch_multiparticle_2011}.  Negativity, $\langle \mathcal{W} \rangle < 0$, indicates entanglement (see Section \ref{sec:entanglement_witnesses}). 

\subsection*{Robust Entanglement Score}
\label{sec:local_equivalence}
Volumetric benchmarks are defined by finite length features ($\mathbf{w} = (w_0,\dots,w_i)$) that then define ensembles of circuits $\mathcal{C}(\mathbf{w})$ and a ``success criterion'' that quantifies performance \cite{blume2020volumetric}. The RES benchmark uses the set of states $\lbrace \mathcal{G}^{\prime}_i \rbrace$ which are equivalent under local operations and classical communication (LOCC) \cite{hein_entanglement_2006}. From an initial $n$-qubit graph state \G{G}{}{} this set of states are generated through sequences of local complement (LC) operations \cite{bouchet1991efficient,bouchet1993recognizing,van2004efficient} (see Section \ref{sec:LC_orbit}).  Thus, from \G{G}{}{}  we use random sequences of LC operations to randomly sample from graph states which are in the same equivalence class. Once the graphs are determined, the corresponding circuits are constructed in two ways: by generating a new graph state circuit
[c.f. Fig.~\ref{fig:graph_state_circuit_construction}(e)] (the \textit{\naive{}} method)
or by appending local unitaries to the state preparation circuit of \G{G}{}{} 
[c.f. Fig.~\ref{fig:graph_state_circuit_construction}(f)] (the \textit{\unitary{}}  method). The \unitary{} ~method maintains the original number of two-qubit operations, whereas for the \naive{}~method the number of two-qubit operations depends on $E(\mathcal{G}^{\prime})$. 

The features used to evaluate the RES benchmark are the quantum circuit width ($w_0=n=|V(\mathcal{G})|$), and graph treewidth $w_1=\mathrm{tw}(\mathcal{G})$ \cite{robertson1986graph,bodlaender1998partial}. Treewidth can be used to identify tree graphs ($\mathrm{tw}(\mathcal{T})=1$ independent of $n$) or complete graphs ($\mathrm{tw}(K_n)=(n-1)$) and is independent of the circuit transpilation. In classical computing, treewidth has been used to quantify problem difficulty  \cite{demaine2005algorithmic,marx2007can}, computational complexity \cite{goles2021impact} and complexity of database queries \cite{grohe2007complexity}.

Success for the RES benchmark is witness function negativity (genuine or biseparable). We assign the RES score for a given backend using the largest set of $n$-qubits that can be genuinely entangled multiplied by the largest treewidth ($\max{w_0} \times \max{\mathrm{w_1}}$). We distinguish the RES scores by the method used: either the \naive{} method (RES-Na\"{i}ve) or the \unitary{} method (RES-Unitary).
\begin{figure*}[htbp]
    \centering
    \includegraphics[width=\textwidth]{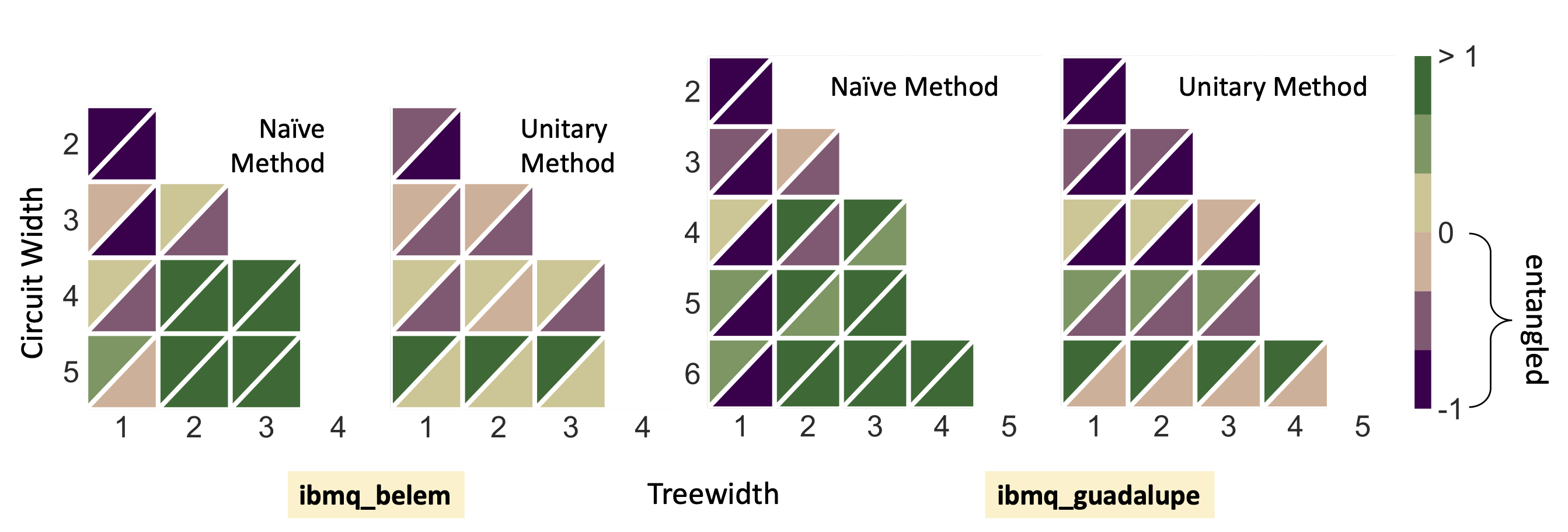}
    \caption{Median genuine witness values aggregated over multiple graph state executions. The upper triangular values are defined by non-mitigated results, the lower triangular values are defined by error mitigated results, using a tensored mitigator defined in Qiskit \texttt{Ignis}. Negative values indicate entanglement. (Left) Benchmarking of \belem{} (RES-N$=3$, RES-U$=6$). (Right) Benchmarking of\guadalupe{} (RES-N$=6$, RES-U$=6$).}
    \label{fig:guadalupe_belem_genuine_heatmaps}
\end{figure*}

\section{Evaluation on IBM hardware}
\label{sec:results}

The RES benchmark was executed on multiple superconducting devices available through IBM Quantum \cite{IBMQ}, listed in Table \ref{tab:ibm_backends} with their respective quantum volumes as reported by IBM. In Figs.~\ref{fig:jakarta_toronto_genuine_heatmaps},\ref{fig:guadalupe_belem_genuine_heatmaps} we present the median genuine witness values evaluated on four IBM backends. Using graph circuits of increasing order we determine what is the largest treewidth, and largest set of qubits ($n$) that can be genuinely entangled and verified. This is not a test of which qubits on the hardware that can be entangled, but starting from arbitrary qubit subsets, what is the largest entangled state that can be prepared and how robust is it to various sources of hardware noise. The smallest number of qubits that we tested entanglement across was $n=2$ and the maximum number of qubits was $n=9$ (or the maximum number of qubits available on a device).  All backends were accessed through cloud-based queues, and the data collection method is described in Appendix \ref{sec:bulk_data}. Tensored mitigators were evaluated during data collection using the Qiskit library \texttt{Ignis} \footnote{Library now depreciated, functionality available in \texttt{qiskit-terra} and \texttt{qiskit-experiments}} \cite{Qiskit_full}. The RES-Na\"{i}ve and RES-Unitary values reported in Table \ref{tab:ibm_backends} are computed from unmitigated values.  Mitigated RES values are reported in Table \ref{tab:ROEM_RES}.
\begin{table}[h]
\begin{tabular}{ |c|c|c|c|c|c| } 
 \hline
 \thead{Backend \\ Name} & Processor & Qubits & \thead{Quantum\\Volume} & \thead{RES-\\Na\"{i}ve} & \thead{RES-\\Unitary}\\ 
 \hline
\texttt{ibmq\_toronto} & r4 & 27 & 32 & 6 & 8 \\ 
\hline
\texttt{ibmq\_guadalupe} & r4p & 16 & 32 & 6 & 6 \\ 
 \hline
\texttt{ibmq\_jakarta} & r5.11H & 7 & 16 & 6  & 12 \\ 
 \hline
\texttt{ibmq\_belem} & r4T & 5 & 16 & 3 & 6 \\ 
 \hline
\end{tabular}
\caption{Quantum backends, processor types, total number of qubits available, published Quantum Volume, RES-Na\"{i}ve and RES-Unitary values.}
\label{tab:ibm_backends}
\end{table}

Preparation of genuine entanglement over arbitrary graphs states represents a stringent benchmark, but biseparable entanglement and sub-chain entanglement are also readily measurable on NISQ devices~\cite{mooney_generation_2021}. In Appendix \ref{appendix:biseparable_witnesses}, we report similar heatmaps for the median biseparable witness results. In general, all backends support high quality bipartite entanglement; the scores are commensurate with other work which has demonstrated multiple two-qubit chains entangled on specific backends.

Overall we observe that in the aggregate it is very difficult to generate and validate genuine $n$-qubit entanglement that is robust against two-qubit gate noise or circuit depth. On the other hand, robust biseparable entanglement was detectable for qubit sizes up to $n=9$.  Readout error mitigation provides moderate improvement for both the genuine and biseparable witness value.  

\section{Effect of Hardware Noise on Stabilizer Measurement}
\label{sec:discussion}
\subsection{Stabilizer Measurement Fidelity}
\label{sec:stabilizer_measurements}
The color scales of the heat maps shown in Figs. \ref{fig:jakarta_toronto_genuine_heatmaps},\ref{fig:guadalupe_belem_genuine_heatmaps} are defined by the median genuine entanglement witness values evaluated over multiple graph states randomly drawn from the LC graph orbit of $n$-qubit hardware graphs. The evaluation of the witness values is dependent on the measurement of individual stabilizers of each graph state, and the ability to generate and detect entanglement in an $n$-qubit graph state is thus determined by the capability of hardware to prepare and measure these stabilizers with high fidelity.  In this section we delve into the individual stabilizer measurements and discuss factors that affect stabilizer measurement fidelity: circuit width, stabilizer weight, circuit scheduling, and treewidth. These features are chosen for our analysis based on the Pearson  r-coefficients (see Appendix \ref{appendix:pearson_correlation} for details).
\begin{figure}[htbp]
  \centering
  \includegraphics[width=\columnwidth]
  {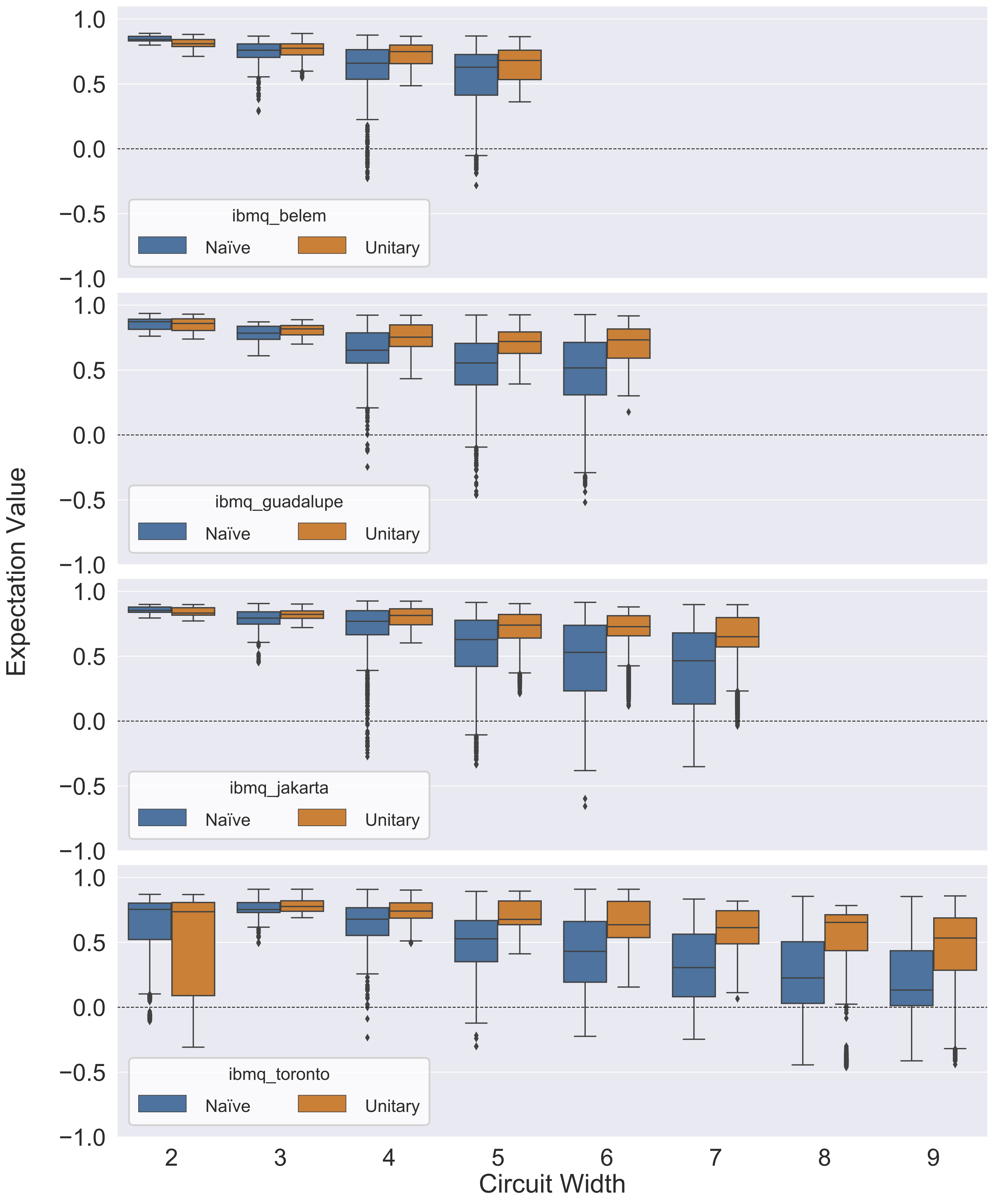}
  \caption{Quartiles of stabilizer expectation values calculated over all measurements evaluated per circuit width.}
  \label{fig:stabilizer_quartiles}
\end{figure} 
As shown in Fig.~\ref{fig:stabilizer_quartiles}, the median stabilizer measurement computed over all $n$ qubit graph states decreases with increasing graph order, with the exception of $n=2$ results generated on \toronto{}. The median expectation values obtained with the \naive{} method of graph state construction decrease more rapidly with the number of qubits than the \unitary{} method. This performance can likely be attributed to the increased number of CNOTs, since an increase in circuit depth through CNOT additions is likely to cause worse performance than an increase in circuit depth caused by local complement (single qubit) operations in the \unitary{} method as single qubit error rates are lower than the CNOT error rates. In Fig. \ref{fig:stabilizer_quartiles} there are notable outliers in the stabilizer expectation values with $n=2$ qubits, specifically for data collection on qubit subsets: $[0,1], [2,3], [3,5]$. On \toronto{} we observe that the $n=2$ stabilizer measurements exhibit a large negative skew -- we discuss the connection to low level calibration metrics in Appendix \ref{appendix:calibration_data}.

\subsection{Error mitigation}
\label{sec:error_mitigation_discussion}
Executing the graph state circuits on near-term quantum devices result in noisy measurements. Hardware noise is generated by many sources such as qubit decoherence, gate noise and readout error.  We consider how individual stabilizer measurements can be affected by two-qubit gate error, readout error or stabilizer weight (non identity measurements). Figure~\ref{fig:jakarta_correlation_matrix}, for example, in Appendix \ref{appendix:pearson_correlation} shows the correlations between these low-level circuit structure datum and expectation values.  The Pearson correlation coefficients are computed pairwise between: circuit width, and number of CNOTs in the transpiled circuits, stabilizer weights, graph treewidth and the observed stabilizer measurement.

The gate scheduling is not optimized, but we investigate the effect of readout error mitigation on benchmark performance.  For our chosen set of states and backends there are many methods available for noise and error mitigation. For the circuits executed on IBM backends we use the error mitigation library \texttt{Ignis} in Qiskit \cite{Qiskit_full} to construct tensored error mitigators: a matrix of noise characterization that is used to post-process the returned distribution over bitstrings prior to evaluation of individual expectation values. Ref \cite{jungnitsch_entanglement_2011} also showed that the inclusion of additional stabilizer measurements improves the robustness of the genuine witness to white noise. Currently, we evaluate these additional terms to test stabilizer measurement stability, but they are not included in the evaluation of the entanglement witness.

\begin{figure}[htbp]
  \centering
  \includegraphics[width=\columnwidth]
  {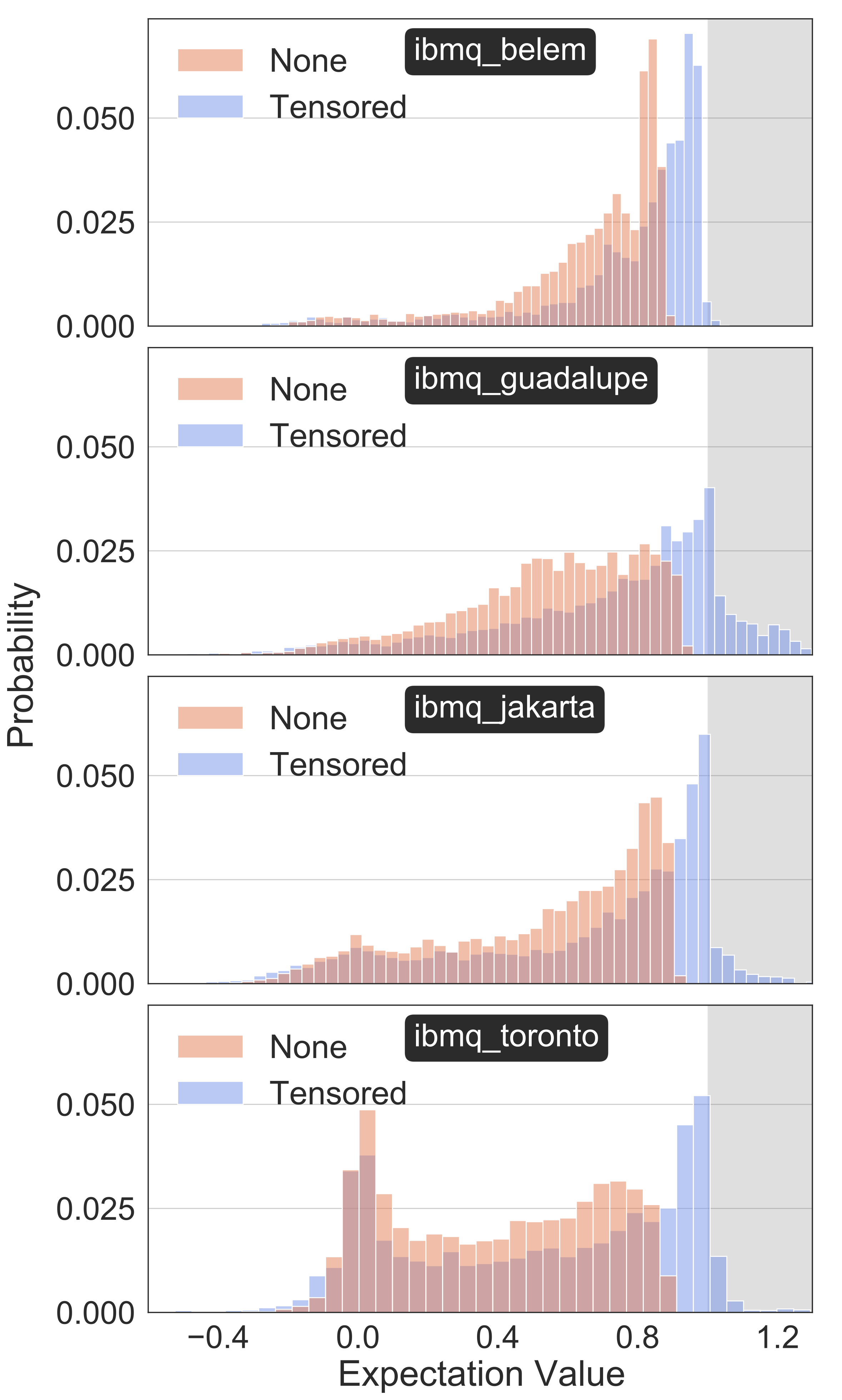}
  \caption{Distribution of stabilizer values measured on SC qubit backends using the \naive{} method.  Readout error mitigation is implemented using tensored construction of assignment error matrices. Shaded region indicates non-physical values generated by noise mitigation.}
  \label{fig:expecation_value_naive_distributions}
\end{figure}

\begin{figure}[htbp]
  \centering
  \includegraphics[width=0.98\columnwidth]
  {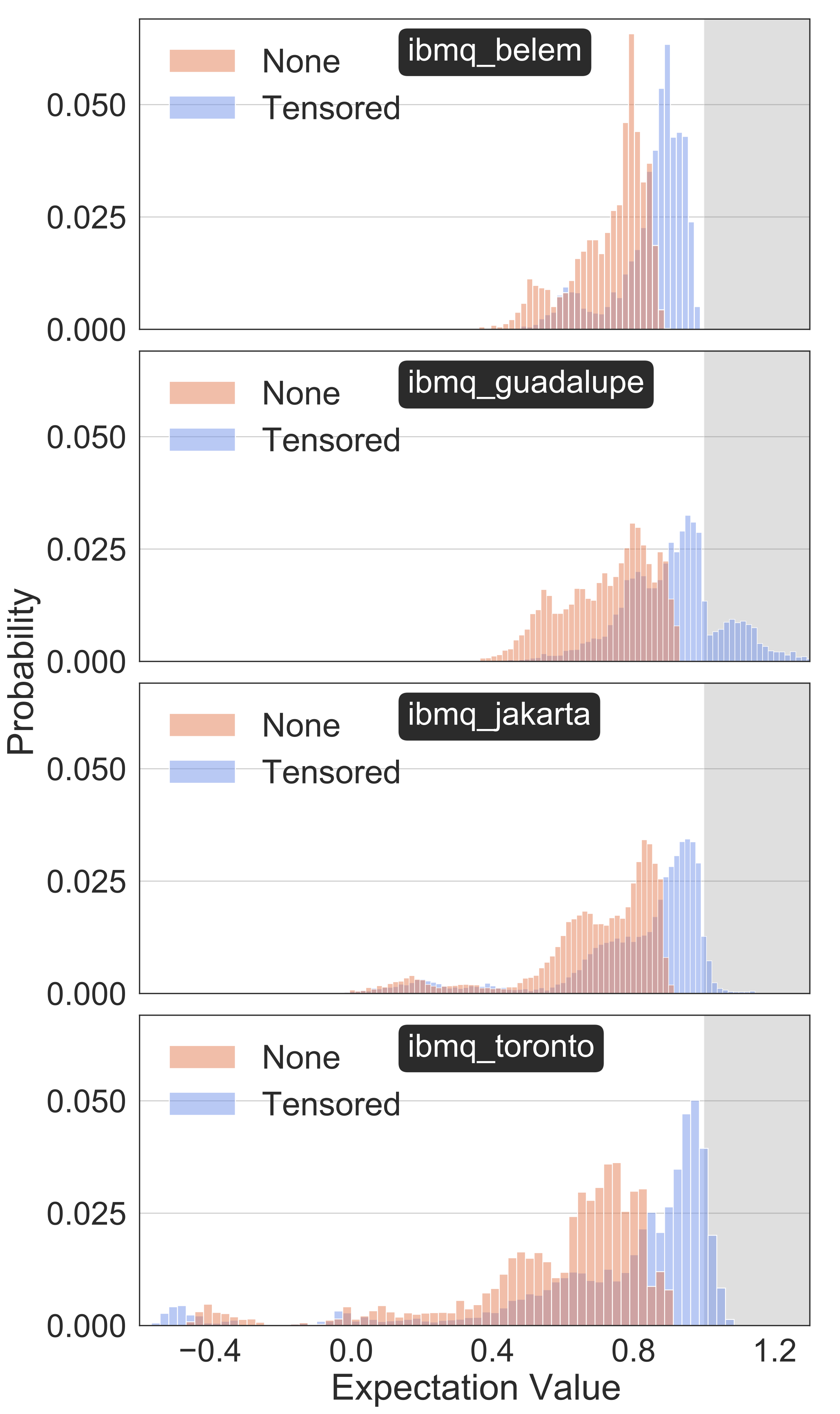}
  \caption{Distribution of stabilizer values measured on SC qubit backends using the \unitary{} method.  Readout error mitigation is implemented using tensored construction of assignment error matrices. Shaded region indicates non-physical values generated by noise mitigation.}
  \label{fig:expecation_value_unitary_distributions}
\end{figure}

In Table \ref{tab:ibm_backends} we reported the RES-N and RES-U scores for each backend, based on the unmitigated median witness values.  As seen in Figs. \ref{fig:expecation_value_naive_distributions} and \ref{fig:expecation_value_unitary_distributions}, the application of readout error mitigation results in a general shift of stabilizer expectation values closer to $1.0$.  As seen in Figs. \ref{fig:guadalupe_belem_genuine_heatmaps}, \ref{fig:jakarta_toronto_genuine_heatmaps} the use of readout error mitigation also improves the median genuine witness value. As a result the RES-N and RES-U scores for each backend are generally improved.  We report the effect of readout error mitigation (ROEM) on the RES scores for each backend in Table \ref{tab:ROEM_RES}.
\begin{table}[h]
\begin{tabular}{ |c|c|c| } 
 \hline
 \thead{Backend \\ Name} & \thead{RES \\ Na\"ive} + ROEM & \thead{RES \\\unitary{}} + ROEM \\ 
 \hline
\texttt{ibmq\_toronto} & 8 & 15\\ 
\hline
\texttt{ibmq\_guadalupe} & 8 & 24 \\ 
 \hline
\texttt{ibmq\_jakarta} & 6 & 15\\ 
 \hline
\texttt{ibmq\_belem} & 6 & 12\\ 
 \hline
\end{tabular}
\caption{Quantum backends, and mitigated RES values evaluated using tensored mitigators.}
\label{tab:ROEM_RES}
\end{table}
In general increased CNOT counts via SWAP additions lead to expected degraded performance. Exceptions exist when readout noise contains strong correlated noise, the stabilizer expectation values veer far from describing entanglement when particular qubits are used which show large amounts of correlated readout noise.
We tested this noise source on SC hardware by examining the effect of readout noise mitigation on the two-qubit subsystem that exhibited the most readout noise on \toronto{}. We found that highly-correlated qubits on this device drastically reduced the probability of successfully preparing an entangled state. 

Figures~\ref{fig:expecation_value_naive_distributions} and~\ref{fig:expecation_value_unitary_distributions} show the change in the distribution of stabilizer expectation values for each backend after a tensored noise mitigator was used to implement readout error mitigation. Overall, the effect of noise mitigation is a general shift rightwards towards the ideal value $\langle g_i \rangle = 1$.  It is possible that the noise mitigation step will result in non-physical values $\langle g_i \rangle > 1$.  When we compute the entanglement witness values with noise mitigated stabilizer measurements we clip the expectation values at $\max{(g_i,1.0)}$.  A future version of the benchmark will include correlated readout noise mitigation rather than the tensored method in order to connect the observed performance with low-level correlations on-chip. Unless explicitly stated our analysis in the remainder of Section \ref{sec:discussion} is of unmitigated measurement data.

\section{Connecting Benchmark Performance to Application Performance}
\label{sec:why}

The focus of this volumetric benchmark is the generation and verification of robust $n$-qubit genuine (multipartite) entanglement. But in benchmarking entanglement we are interested in exploring the control of long-range interactions and correlations between qubits \cite{streltsov2017colloquium}. In contrast to existing volumetric benchmarks which rely on random circuits, we argue that our benchmark can be closely related to near-term application performance or expanded and adapted to other applications.  For example, connections between instantaneous quantum polynomial (IQP) circuits and state preparation for weighted graph states \cite{hayashi2019verifying}, or planar graph states \cite{fujii2017commuting} have been established.  We can apply our benchmark to planar and non-planar graph states, the circuit construction can be expanded to include hypergraphs and weighted graphs, although state verification will need to be modified away from stabilizer measurements~\cite{hein_entanglement_2006}. In this section we discuss additional areas where benchmark performance can be used to infer future application performance on NISQ hardware. 

First, the metric acts as a surrogate model for near-term algorithm and application performance. Many applications executed on NISQ hardware may not map directly into the sparse qubit connectivity. When a densely connected circuit on $n$ qubits is transpiled into a sparsely connected device, the resulting instruction set may have many additional two-qubit gates. Using ideal graphs, i.e. those that embed into the native hardware graph with minimal overhead, we can estimate a ``best case'' performance in the presence of hardware noise (gate noise, qubit decoherence and readout error).  The non-ideal graph states from the \unitary{} method then test the robustness of n-qubit entanglement to swap gate noise or circuit depth.  

Second, the benchmark gathers data through the construction of various graph states, and the measurement of $n$ stabilizers through repeated state preparation.  This data can be leveraged to develop noise mitigation methods for either the bitstring distribution or the stabilizer expectation values.  Additionally, the variability in the individual stabilizer values can be used to infer the stability of state preparation for variational algorithms \cite{cerezo2021variational}.  Overall the ability to efficiently characterize how close a prepared state is to a target graph state, without the need for full tomography, makes this benchmark appealing and scalable. 

\section{Methods}
\label{sec:methods}

Graph states provide us a flexible and extensible means of generating entangled states which also provides insight into the hardware capabilities \cite{toth_detecting_2005,toth_entanglement_2005,jungnitsch_entanglement_2011,hein_entanglement_2006}. This section will describe the theory behind entanglement witnesses, the design of the NISQ circuits for witness evaluation and finally how the metric is evaluated.

\subsection{Benchmark circuit construction}
\label{sec:graph_state_circuits}
A given NISQ device is characterized by the qubit technology, the hardware noise, and the hardware native graph \G{B}{} which  is the fixed layout of couplers and waveguides between physical qubits.  This layout defines which two-qubit gates can be implemented with low overhead, or without incurring additional swap gates.  For each of the IBM backends we use in this work (\belem{}, \jakarta{}, \guadalupe{}, and \toronto{}), these layouts are given in Fig. \ref{fig:all_machine_layouts} of the Appendix. 

The state preparation circuit for \Gstate{G}{n}{E} is constructed in two steps [c.f. Fig.~\ref{fig:graph_state_circuit_construction}(c)]. First, the all-zero register $|0\rangle^{\otimes n}$ is transformed into $|+\rangle^{\otimes n}$ by applying Hadamard gates to all qubits in the register.  Next, each two-qubit entanglement operation defined by the undirected edge set $E(\mathcal{G})$is decomposed into a CNOT gate and two Hadamard gates acting on the target qubit following the sequence used in \cite{wang_16qubit_2018}. 
In this initial demonstration of the benchmark, we limit the degree of optimization in gate scheduling used during the transpilation step.  The graph state construction does not schedule gates to reduce gate depth or optimize gate scheduling in general. We apply barriers to prevent the consolidation of the single qubit gates applied in the \unitary{} method.  In future demonstrations, different compilation and transpilation techniques could be incorporated to improve the benchmark performance.

\subsection{LC equivalent state preparation}
\label{sec:LC_orbit}
\begin{figure}[htbp]
  \centering
  \includegraphics[width=\columnwidth]{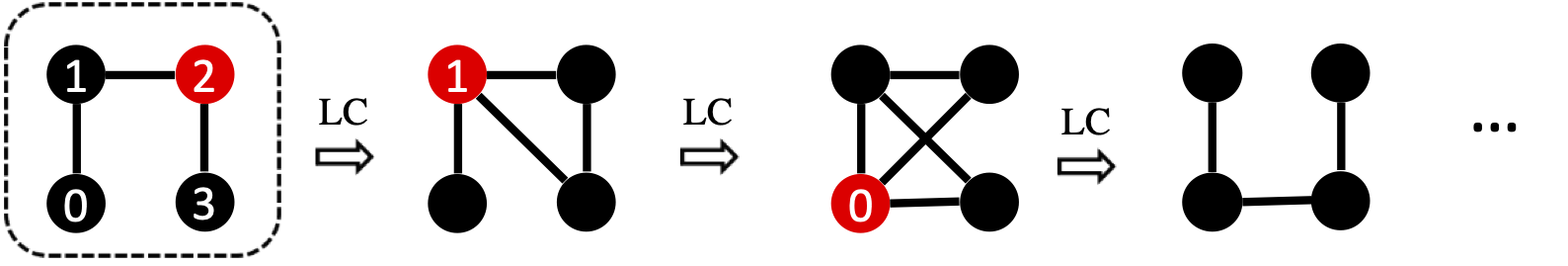}
  \caption{The transformation of an initial 4 vertex path graph (far left) using local graph complements.  The vertex used to define the local complement at each step is highlighted in red.}
  \label{fig:sample_LC_sequence}
\end{figure} 
Evaluating the entanglement metric samples from a finite set of locally equivalent graph states for an initial graph state.  This equivalence class is defined by local complement operations (LC). A local complement of a graph \G{G}{} is defined by a vertex $v_i \in V(\mathcal{G})$ and its neighborhood $N(v_i)$. The complement of the neighborhood graph $N^{\prime}(v_i)$ is used to define $\mathcal{G}^{\prime} = \mathcal{G} \bigcup N^{\prime}$.  An example of LC operation on a 4 vertex path graph is shown in Fig. (\ref{fig:sample_LC_sequence}).  We describe the two methods for implementing LC operations on graph states that we evaluate in this study:  the \unitary{} method and the \naive{} method considering an initial graph \G{G}{} and a specific LC operation ($\ell$).

The \unitary{} method implements the action of a LC operation ($\ell$) on vertex ($a$) using local single qubit gates \cite{hein_entanglement_2006,PhysRevA.83.042314},
\begin{equation}
    \ell_{a} = e^{-i\frac{\pi}{4}\sigma_x^{a}}\prod_{b \to a} e^{i\frac{\pi}{4}\sigma_z^{b}}
    \label{eq:generator_transform}
\end{equation}
where $b\to a$ denotes all neighbors of vertex (a). The graph state circuit is initially constructed with the edge list of \G{G}{}, then the LC operation ($\ell: \mathcal{G} \to \mathcal{G}^\prime$) modifies the circuit $\mathcal{U}_\mathcal{G} \to \mathcal{U}^{\prime}_{\mathcal{G}^\prime}$ by appending X- and- Z axis rotations to $\mathcal{U}_\mathcal{G}$ according to Eq. \ref{eq:generator_transform}. The \unitary{} method also modifies the stabilizer strings $\mathcal{M}_{\mathcal{S}} \to \mathcal{M}_{\mathrm{LC}(\mathcal{S})}$ \cite{cabello2009compact,hein_entanglement_2006}. This method keeps the number of controlled rotation gates minimal, at the expense of circuit depth and with the requirement of  different basis transformations ($X, Y, Z$).

The \naive{} method implements a LC operation by wholly redefining \G{U}{G}.  The LC operation is first applied to the underlying graph, $\ell: \mathcal{G} \to \mathcal{G}^\prime$, and the resulting graph $\mathcal{G}^{\prime},$ is used to construct the graph state circuit.  This method of modifying the underlying graph may lead to graph state circuits with two qubit gates that do not embed into \G{B}{} with minimal overhead, as a result this method may incur additional noise and overhead in the form of swap gates.  However, the definition of the stabilizer strings will only using the measurement settings $X, Z$.

We close this section by noting that each subgraph extracted from the hardware native graphs corresponded to graphs in different LC equivalence classes as shown in Table V of \cite{hein_entanglement_2006}. This is relevant because while different graphs can be contained as subgraphs of individual graphs-- for example the $4$-vertex star graph is seen as a subgraph in several graphs of the $4$-vertex path graph orbit shown in Fig. \ref{fig:sample_LC_sequence},  the graphs generated in each respective LC orbit were not contained in another LC orbit.

\subsection{Entanglement witnesses for graph states}
\label{sec:entanglement_witnesses}
A state is entangled when it is not fully separable. Multipartite entanglement has richer structure than a bipartite entanglement. A state that is not biseparable is called genuinely multipartite entangled. Biseparable entanglement witnesses and genuine entanglement witnesses can be used to characterize the entanglement. Constructing these witnesses for a general state $|\Psi\rangle$  is difficult, but since we are considering graph states we can leverage the graph state stabilizers. 

For each vertex in $v_k \in V(\scriptG{G})$ there is an associated generator constructed from the Pauli $Z, X$ operators,
\begin{equation}
 g_k^{(\scriptG{G})} = X^{(k)} \prod_{l \to k} Z^{(l)},
 \label{eq:generator_def}
\end{equation}
where $l\to k$ denotes the product over all neighbors of vertex (k). For vertices that are not neighbors of (k) an identity operator is inserted.  Each $n$-qubit graph state has a set of $n+1$ stabilizers, the $n$ generators plus the identity string. These Pauli strings are measured and their expectation values are used to compute the entanglement witness functions. These measurements are done serially with multiple preparations of the graph state (see Fig. \ref{fig:graph_state_circuit_construction}(b)) and measuring each individual stabilizer string.  

The genuine entanglement witness operator $\mathcal{W}_{\mathrm{G}}$ can be constructed using all $n$ generator measurements: 
\begin{equation}
 \mathcal{W}_{\mathrm{G}} = (n-1)\mathbbm{1} - \sum_k g_k^{\scriptG{G}}.   
 \label{eq:gen_ent_wit}
\end{equation}
Each genuine entanglement witnesses constructed as in Eq. \ref{eq:gen_ent_wit} requires $n$ stabilizer measurements for each $n$-qubit circuit. This implementation is not optimized for noise robustness: as stated in Ref. \cite{toth_detecting_2005} the witnesses defined with $n$ stabilizer measurements are only robust against noise levels up to $p_{noise}<1/n$. Many studies have investigated entanglement witness construction ~\cite{chruscinski_entanglement_2014, hauke_measuring_2016, han_construction_2016, shahandeh_ultrafine_2017, waegell_benchmarks_2019, amaro_design_2020, dirkse_witnessing_2020, roik_accuracy_2021, rodriguez-blanco_efficient_2021}. Entanglement witnesses were also recently used in a fault-tolerant weight-4 parity check measurement scheme~\cite{hilder2021fault}, which demonstrated genuine six-qubit multipartite entanglement in their shuttling-based trapped-ion quantum computer. 
However, for the purposes of our benchmark, we consider the witness construction in Eqs. \ref{eq:gen_ent_wit},\ref{eq:bisep_ent_wit} to be optimal because it uses the same number of stabilizer measurements per graph, and the entanglement of each generated graph state is evaluated using witnesses with the same level of noise robustness.  

\subsection{Data Collection}
\label{sec:bulk_data}
Given $n$ (number of qubits), we can induce a subgraph of $n$ vertices on a backend with hardware native graph $\mathcal{G}_\mathcal{B}$ (see Fig. \ref{fig:all_machine_layouts} in Appendix \ref{appendix:GHZ_states}). This subgraph can be used to prepare a specific graph state, and we sample a set of LC equivalent graph states from its LC graph orbit. This is done by generating random sequences of LC operations: $[a,b,c]$ corresponds to applying the LC at vertex (a), followed by the LC at vertex (b), followed by the LC at vertex (c).  Depending on the induced subgraph, the LC graph orbit can be very small (e.g. see the LC graph orbits associated with GHZ states in Appendix \ref{appendix:GHZ_states}) or it can contain a large number of graphs (see Table V of Ref. \cite{hein_entanglement_2006}).  For $n$ qubits we generate a minimum of $2^{n+1}$ sequence of LC sequences.  Each sequence length is drawn uniformly at random from $[1,2n]$ and the only constraints applied to the random sequences is that duplicate operations (e.g. $[a,a]$) are consolidated to $[a]$.  We do not require that unique graphs are generated by each sequence, it is possible that two sequences $\ell = [a,b,c]$ and $\ell^{\prime} = [a^{\prime},b^{\prime}, c^{\prime}]$ will generate the same graph. It is also possible that a sequence will have no net effect on the induced subgraph of the hardware native graph.  

Once the LC sequences have been generated, each individual graph state preparation circuit is constructed in Qiskit $n$ times to evaluate the $n$ stabilizers of the target graph state, plus the identity string (measurement in the computational basis only).  Overall, a sequence of $L$ LC sequence of a $n$ qubit graph state will require $nL$ circuits. This data is collected using large job batches.  Each backend available via the IBM Q Experience has a maximum number of experiments that can be sent under the same job batch.  We generate circuit sets (see pseudocode in Appendix \ref{appendix:pseduocode}) which are partitioned into job batches (each graph, each set of stabilizers) such that the maximum number of experiments are sent to the backend. However, if multiple batches are needed, then the jobs are partitioned such that all circuits for a specific graph state are sent in the same batch. After a batch has been sent to a backend and executed, we send the necessary circuits to evaluate the tensored mitigator.
\section{Data Availability}
\label{sec:data_availability}
The datasets generated during and/or analyzed during the current study are available from the corresponding author on reasonable request.

\section{Conclusions}
\label{sec:conclusions}

The benchmark we have introduced in this paper has applications and benefits to near-term, NISQ computing as well as longer term applications. It is challenging to develop metrics that capture the impact of errors, but including additional states of the Hilbert space would give a more thorough characterization of the device. In the near term, the data collected during the execution of this study has amassed a large corpus of state preparation results. This resource will be leveraged to advance near-term noise mitigation methods that can be applied to the $2^n$ distribution of bitstring counts or the individual expectation values. Also in the near term, our benchmark can be used to infer performance on variational quantum algorithms (VQAs). Our benchmark relies on Clifford circuits, and to demonstrate the potential benefits for VQAs we point to recent studies that have leveraged Clifford circuits for bootstrapped initialization of VQAs \cite{ravi2022cafqa}. In this work, Clifford circuits are used to find regions of the optimization landscape to facilitate training.  The specific graph state preparations used in our benchmark can capture the dynamics and impact of errors in different regions of the Hilbert space, and in different bases. 

Adaptable, informative benchmarking will assist in the development of hardware and algorithms that reach this milestone. While the generation of $n$-qubit multi-partite entanglement may not be a final goal of near-term computing, quantifying this entanglement can be connected with many applications and serves as an effective way to benchmark quantum hardware.

\section*{Competing Interests}
The Authors declare no Competing Financial or Non-Financial Interests

\section*{Author Contributions}
K.E.H., N. L., S. E., A. K., A. F. , K.Y.A., and R. C. P. contributed to the scientific design.\\
K.E.H., A. K., A. F. , K.Y.A., and R. C. P. contributed to the data collection.\\
N. L., K.E.H., G. B., A.F., A.K., T. M., H. C. and M. K. contributed to the software development and testing.\\
K.E.H., A. F., A.K., R.C.P. , contributed to the data analysis and interpretation.  \\
All authors contributed to the manuscript writing and editing.

\section*{Acknowledgements}
The authors would like to thank Michael R. Geller for helpful discussions about multipartite entanglement during the design of this benchmark. 

The scientific design, data collection, manuscript writing and data analysis was supported in part as part of the ASCR Testbed Pathfinder Program at Oak Ridge National Laboratory under FWP ERKJ332 (K.E.H., T. M., K.Y.A., A.F., and R. C. P.). The software development and testing  (H. C. and M. K.) was supported in part by the U.S. Department of Energy, Office of Science, Office of Workforce Development for Teachers and Scientists (WDTS) under the Science Undergraduate Laboratory Internship program. 
A.F. and A.F.K were supported by the National Science Foundation under Grant No. NSF DMR-1752713 (for planning, formal development, and software development) and by the ASCR Testbed Pathfinder Program (data interpretation and manuscript writing). 
K.Y.A. was supported by MITRE Corporation TechHire and Quantum Horizon programs during data collection and manuscript writing. \copyright 2022 The MITRE Corporation. ALL RIGHTS RESERVED. Approved for public release. Distribution unlimited 21-03848-5. 

This research used quantum computing system resources of the Oak Ridge Leadership Computing Facility, which is a DOE Office of Science User Facility supported under Contract DE-AC05-00OR22725. Oak Ridge National Laboratory manages access to the IBM Q System as part of the IBM Q Network.

\bibliographystyle{unsrt}  

\appendix
\label{appendix}

\section{Biseparable Witnesses}
\label{appendix:biseparable_witnesses}
\begin{figure}[htbp]
    \centering
    \includegraphics[width=\columnwidth]{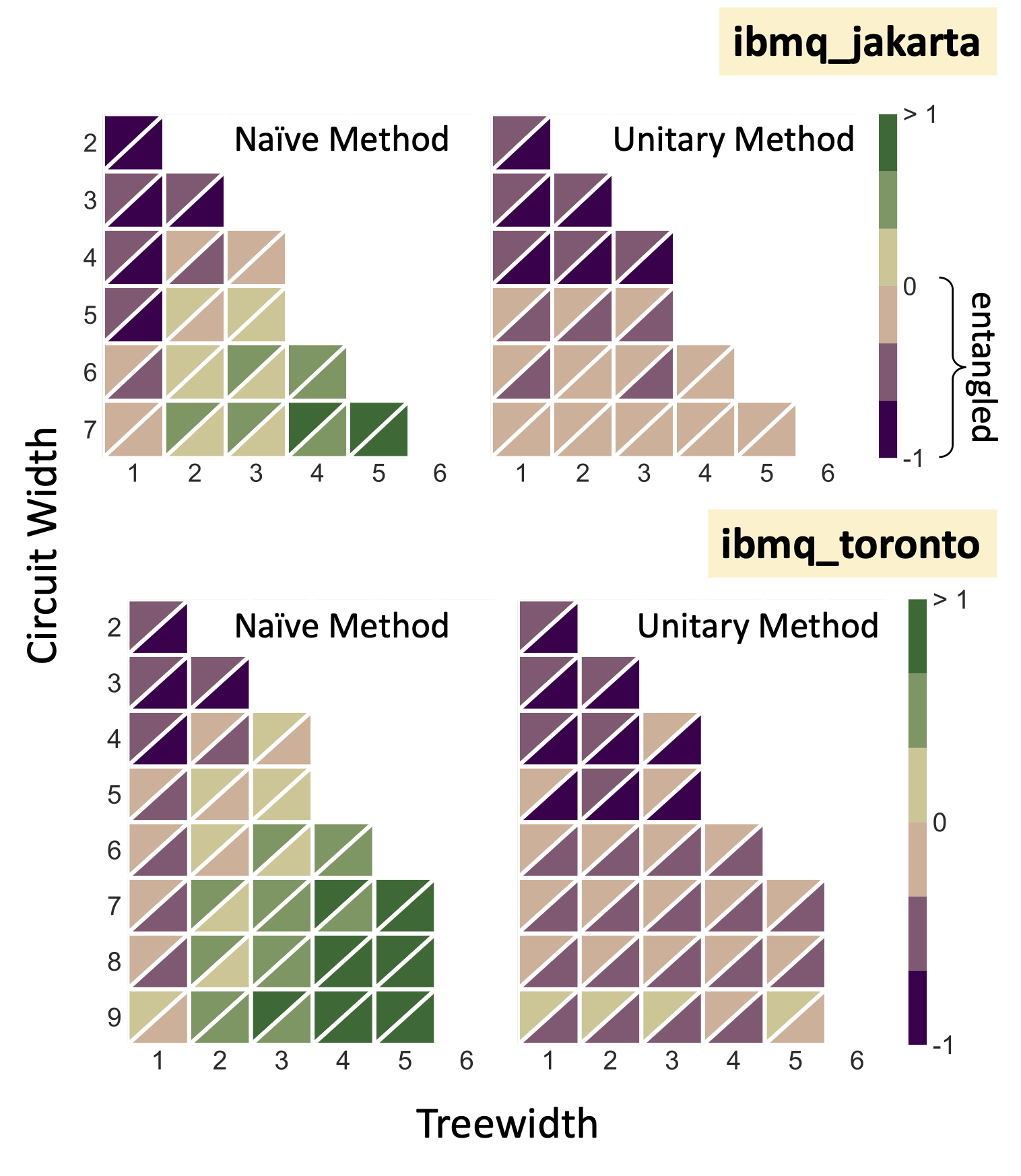}
    \caption{Median biseparable witness values aggregated over multiple graph state executions. The upper triangular values are defined by non-mitigated results, the lower triangular values are defined by error mitigated results, using a tensored mitigator defined in Qiskit Ignis. Negative values indicate that entanglement can be generated and detected. (Top) Volumetric benchmarking of \jakarta{} (Bottom) \toronto{} using biseparable entanglement witnesses. }
    \label{fig:bisep_entanglement_heatmaps_jakarta_toronto}
\end{figure}
\begin{figure}[htbp]
    \centering
    \includegraphics[width=\columnwidth]{ 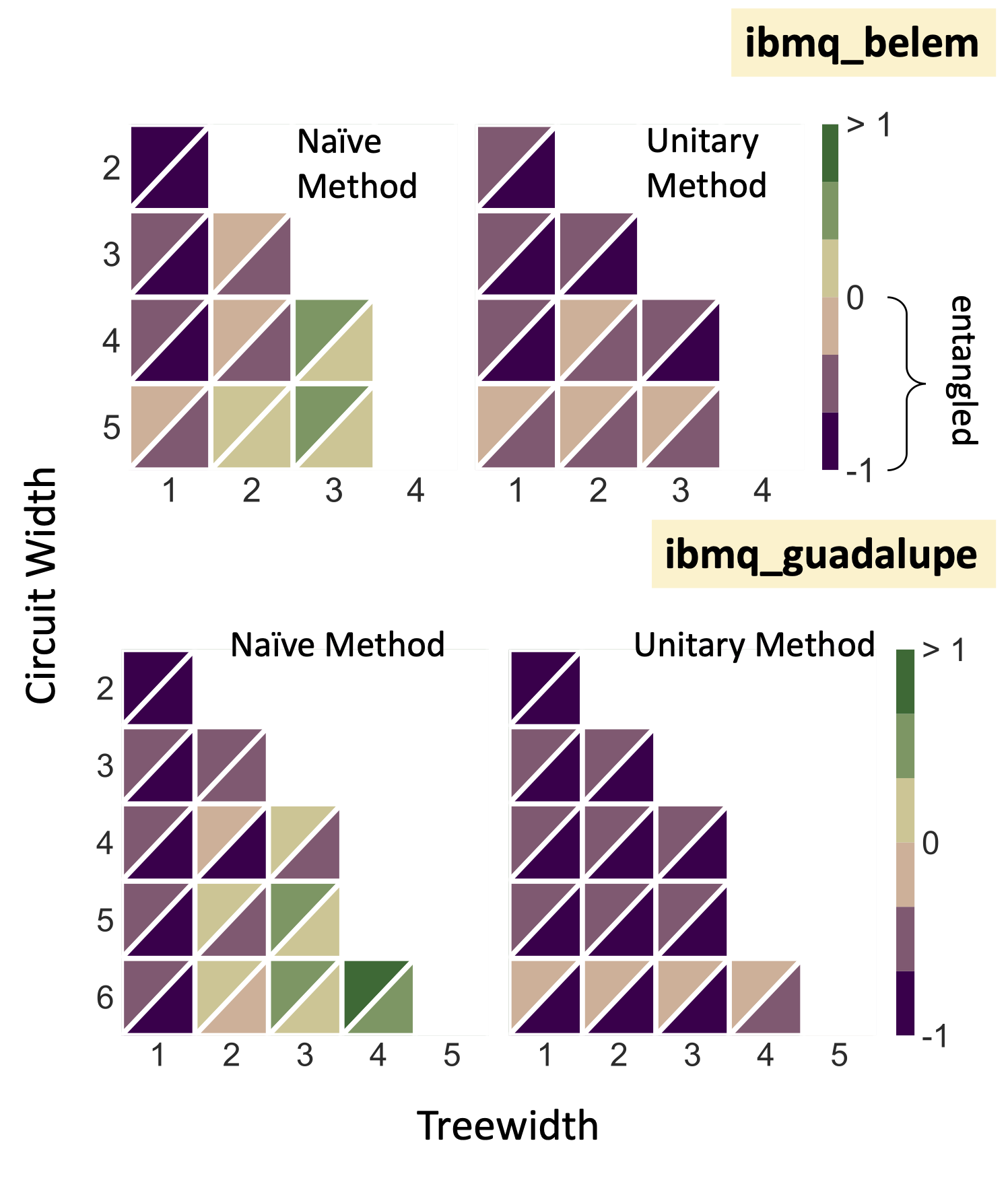}
    \caption{Median biseparable witness values aggregated over multiple graph state executions. The upper triangular values are defined by non-mitigated results, the lower triangular values are defined by error mitigated results, using a tensored mitigator defined in Qiskit Ignis. Negative values indicate that entanglement can be generated and detected. (Top) Volumetric benchmarking of \belem{} (Bottom) \guadalupe{} using biseparable entanglement witnesses. }
    \label{fig:bisep_entanglement_heatmaps_guadalupe_belem}
\end{figure}

Evaluating the biseparable entanglement witnesses quantifies the separability between pairs of qubits in the $n$-qubit state. A biseparable entanglement witness can be constructed for each edge $e_{ij} \in E(\scriptG{G})$ and only requires two measurements (see \cite{toth_entanglement_2005})
\begin{equation}
    W_{i,j}(G)=\mathbbm{1} - g_i^{(\scriptG{G})} - g_j^{(\scriptG{G})}.
\label{eq:bisep_ent_wit}
\end{equation}
In Figures \ref{fig:bisep_entanglement_heatmaps_jakarta_toronto}, \ref{fig:bisep_entanglement_heatmaps_guadalupe_belem} we similarly plot the median value aggregated over all biseparable witness values evaluated from the LC graph orbits.  The biseparable witnesses are evaluated using the same set of stabilizer measurements collected from graphs sampled from the LC graph orbit.

\section{GHZ state preparation}
\label{appendix:GHZ_states}
\begin{figure}[htbp]
  \centering
  \includegraphics[width=\columnwidth]{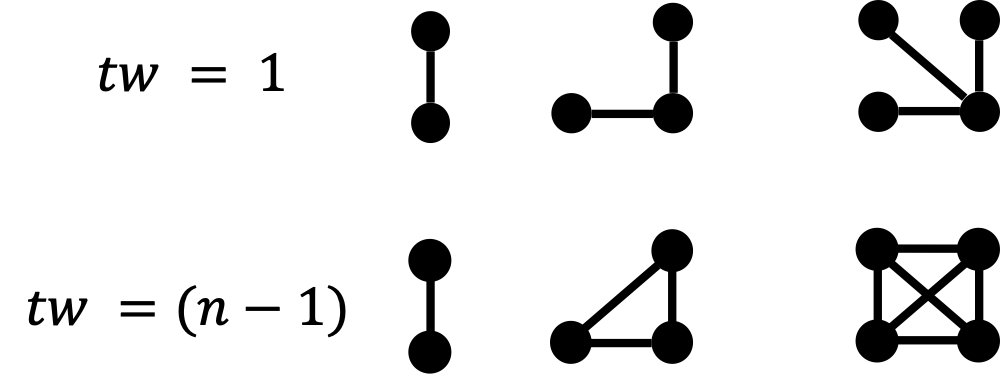}
  \caption{The graph orbits associated with the $|\mathrm{GHZ}_2\rangle,|\mathrm{GHZ}_3\rangle,|\mathrm{GHZ}_4\rangle$ states that can be prepared from subgraphs directly extracted from the $4$ IBM backends we have discusssed in this work. The graphs in each orbit have treewidths of either $1$ or $n-1$. }
  \label{fig:GHZ_orbit}
\end{figure} 
\begin{figure*}[htbp]
    \centering
    \begin{subfigure}[t]{0.3\textwidth}
        \centering
    \includegraphics[height=1.in]{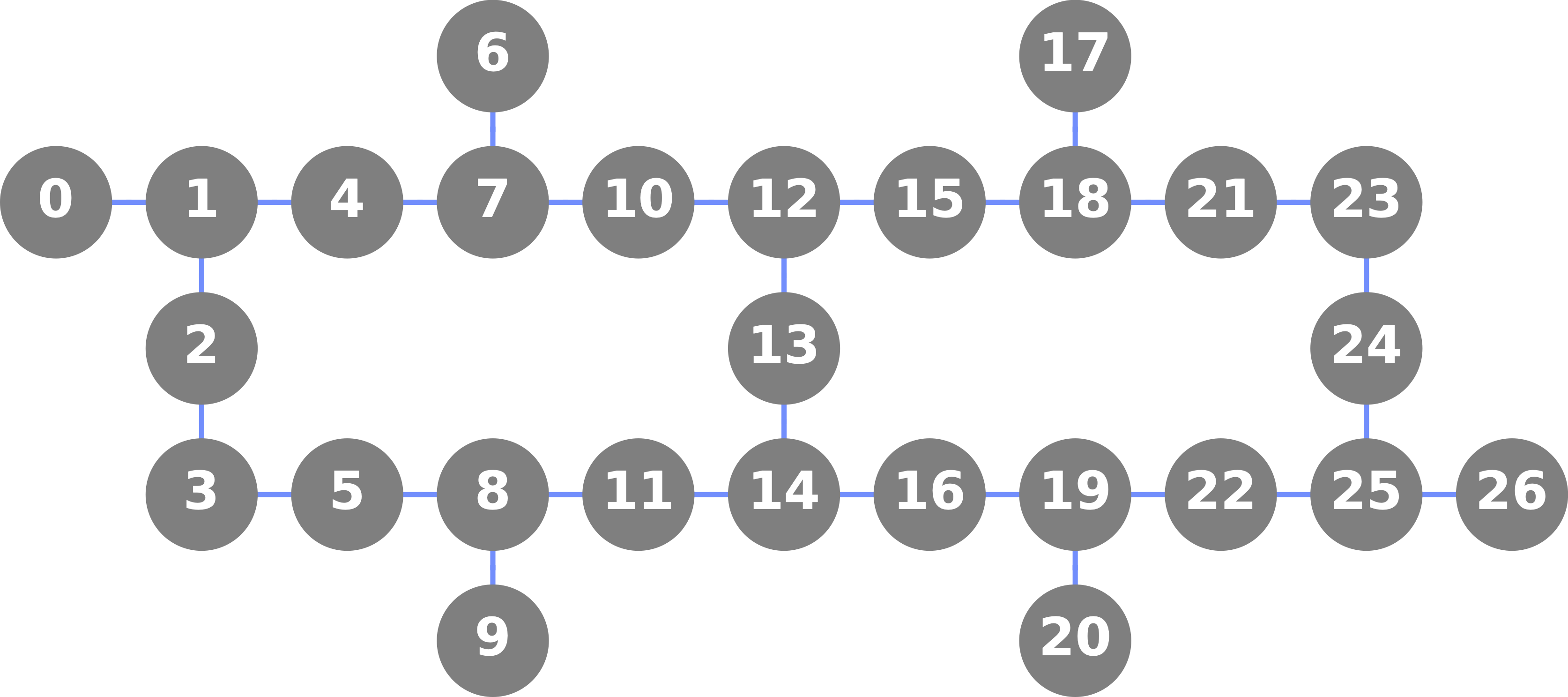}
    \end{subfigure}%
    \hspace{3em}%
    \begin{subfigure}[t]{0.25\textwidth}
    \centering
         \includegraphics[height=1.in]{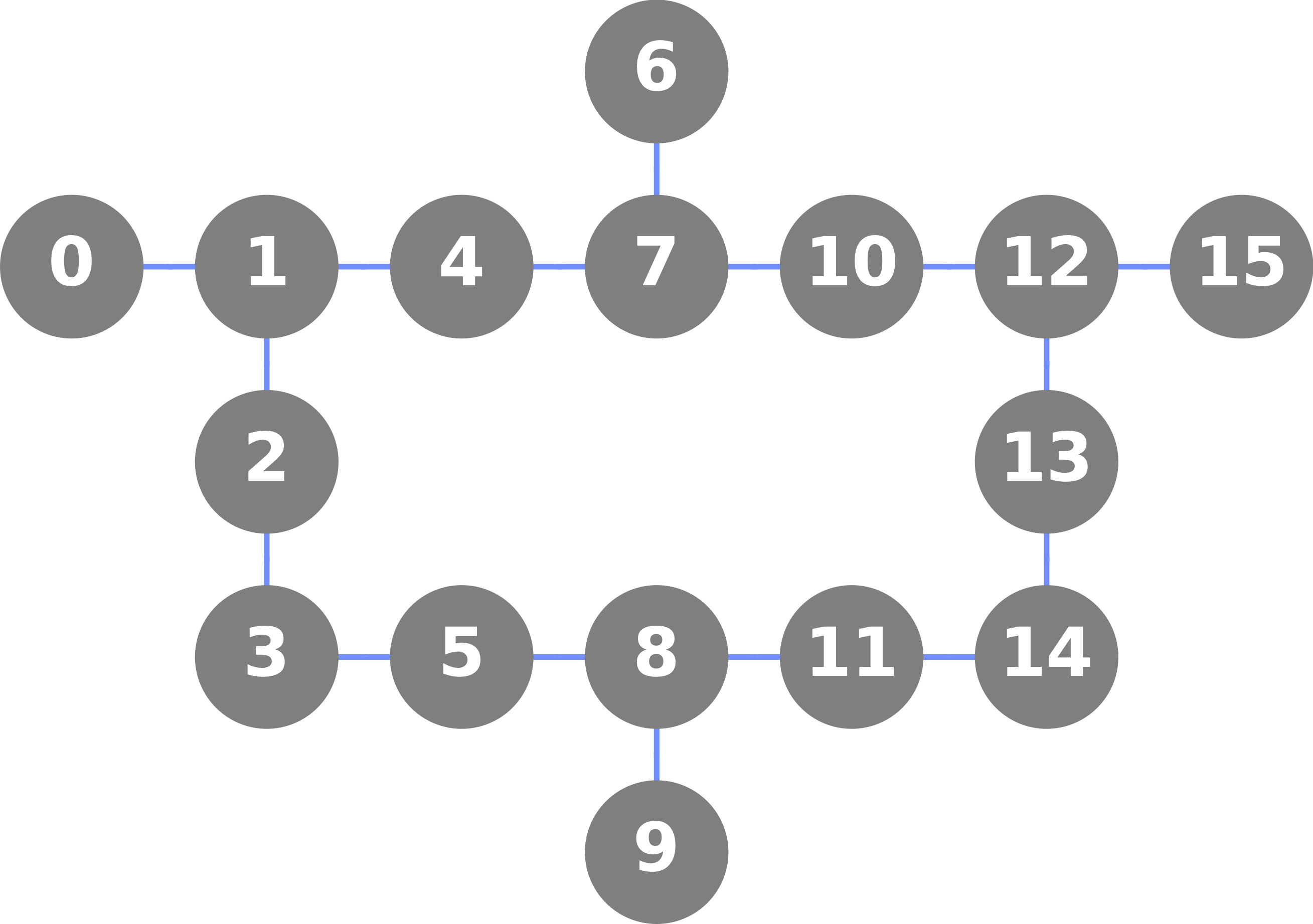}
        \end{subfigure}
    \begin{subfigure}[t]{0.15\textwidth}
        \centering
        \includegraphics[height=0.8in]{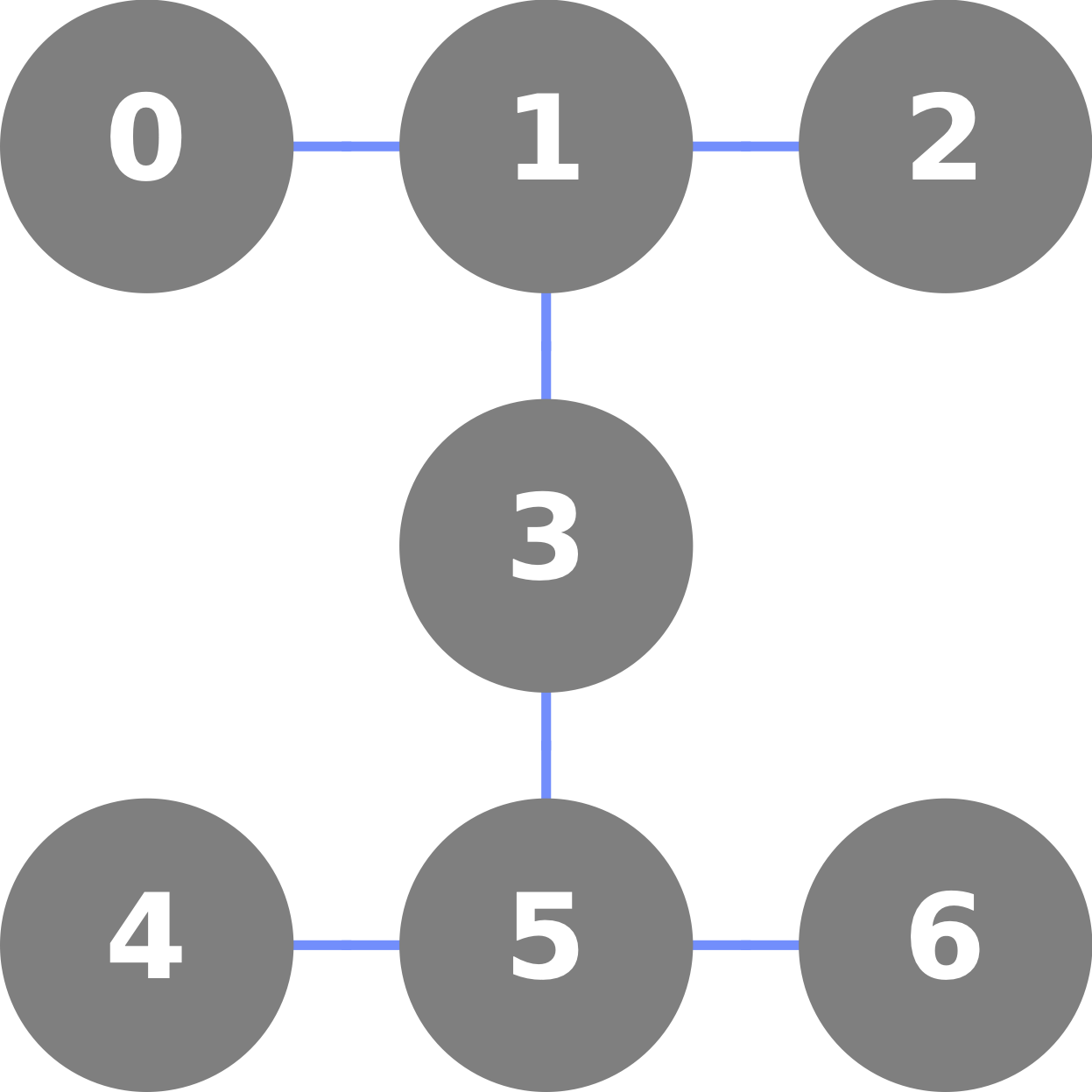}
    \end{subfigure}%
    \hspace{2em}%
    \begin{subfigure}[t]{0.15\textwidth}
        \centering
      \includegraphics[height=0.8in]{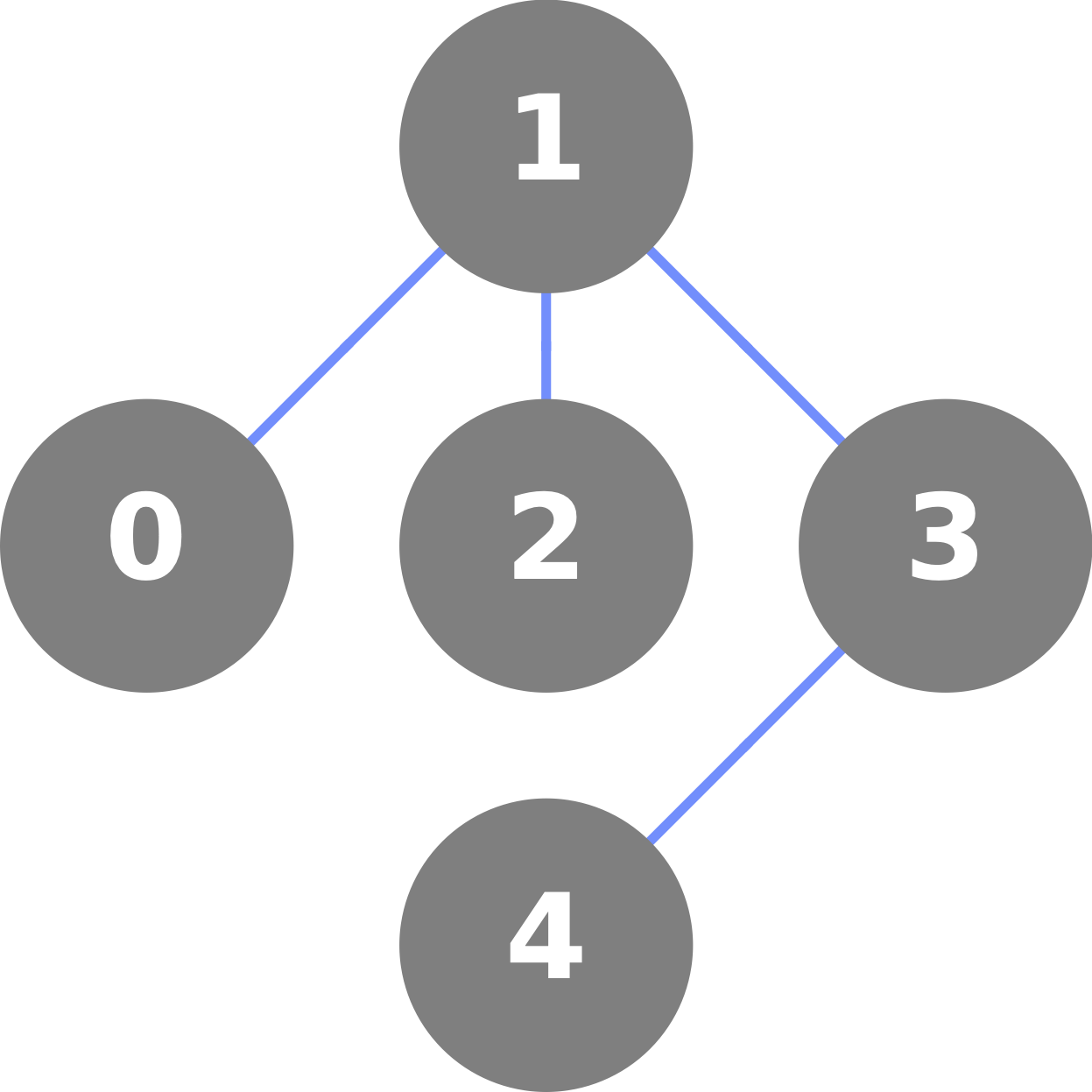}
    \end{subfigure}
        \caption{(Left to Right) Hardware native graphs of \toronto{}, \guadalupe{}, \jakarta{}, and \belem{}. On \toronto{}, the largest qubit subset used in this study contained 9 qubits.  On \guadalupe{}, the largest qubit subset used in this study contained 6 qubits. On \jakarta{} and \belem{} graph states were generated which used all qubits on the backend.}
       \label{fig:all_machine_layouts}
\end{figure*}
\par We have evaluated the robustness of entanglement generated on four IBM machines using our benchmark. Each backend has a specific hardware native graph which defines the possible set of subgraphs that can be extracted on $n$ qubits.  By design, our benchmark only considers the graph orbits associated with induced subgraphs of the hardware native graphs defined for a specific backend. In this section we highlight the optimal preparation of $n$-qubit GHZ state $|\mathrm{GHZ}_n\rangle$ encountered during the evaluation of our metric.  For the qubit layouts shown in Fig.~\ref{fig:all_machine_layouts}, each backend can prepare the $|\mathrm{GHZ}_2\rangle,|\mathrm{GHZ}_3\rangle,|\mathrm{GHZ}_4\rangle$ graph states from hardware subgraphs (see Fig. \ref{fig:GHZ_orbit}). The LC graph orbit associated with a $n$-vertex star graph contains only the star graph ($tw=1$) and the $n$-vertex complete graph ($tw=n-1$) (see Fig. \ref{fig:GHZ_orbit}). 

In Tables \ref{tab:GHZ_genuine_only} and \ref{tab:GHZ_bisep_only} we highlight the lowest observed genuine entanglement witnesses and the lowest observed biseparable entanglemnt witnesses for GHZ graph states and their orbits.  
\begin{table}[h!]
\centering
\begin{tabular}{ |p{2.8cm}|c|c|c|c| } 
\hline
\multicolumn{5}{|c|}{Genuine Witness-GHZ orbit} \\ 
\hline
 Name & Qubits & Star & Complete & Method \\ 
 \hline
 \multirow{6}{*}{\texttt{ibmq\_belem}} & 2 & -0.740234 & -0.740234 & \naive{} \\
  \cline{2-5}
 & 2 & -0.728271 & -0.728271 & \unitary{} \\
  \cline{2-5}
 \cline{2-5}
 & 3 &  -0.458252 & -0.116211 & \naive{}\\ 
  \cline{2-5}
& 3 &  -0.541260 & -0.458252 & \unitary{}\\ 
 \cline{2-5}
  \cline{2-5}
  & 4 &  -0.127197 & 1.232422 & \naive{}\\ 
   \cline{2-5}
& 4 & -0.135742 & -0.130615 & \unitary{}\\ 
 \hline
 \multirow{6}{*}{\texttt{ibmq\_toronto}} 
 & 2 & -0.718018 & -0.718018 & \naive{} \\ 
  \cline{2-5}
 & 2 & -0.710693 & -0.710693 & \unitary{} \\ 
  \cline{2-5}
 \cline{2-5}
 & 3 &  -0.476318 & -0.185791 & \naive{} \\ 
  \cline{2-5}
& 3 &  -0.471680 & -0.459717 & \unitary{} \\ 
 \cline{2-5}
 \cline{2-5}
& 4 &  0.143799 & 1.496826 & \naive{} \\ 
 \cline{2-5}
& 4 &  0.142578 & 0.180176 & \unitary{} \\ 
\hline
 \multirow{6}{*}{\texttt{ibmq\_guadalupe}} 
 & 2 & -0.860352 & -0.860352 & \naive{} \\ 
  \cline{2-5}
 & 2 & -0.856201 & -0.856201 & \unitary{} \\ 
  \cline{2-5}
 \cline{2-5}
& 3 &  -0.568359 & -0.191162 & \naive{}\\ 
 \cline{2-5}
 & 3 & -0.587158 & -0.565186 & \unitary{} \\ 
  \cline{2-5}
 \cline{2-5}
& 4 &  -0.250244 & 0.914551 & \naive{}\\ 
 \cline{2-5}
 & 4 & -0.104980 & -0.032227 & \unitary{} \\ 
 \hline
 \multirow{6}{*}{\texttt{ibmq\_jakarta}} 
 & 2 & -0.792969 & -0.792969 & \naive{} \\ 
  \cline{2-5}
 & 2 & -0.779785 & -0.779785 & \unitary{} \\ 
  \cline{2-5}
 \cline{2-5}
& 3 &-0.640381 & -0.321045 & \naive{} \\ 
 \cline{2-5}
 & 3 & -0.552246 & -0.540039 & \unitary{} \\ 
  \cline{2-5}
 \cline{2-5}
& 4 & -0.447998 &  0.727783 & \naive{} \\ 
 \cline{2-5}
 & 4 & -0.422363 & -0.377197 & \unitary{} \\ 
 \hline
\end{tabular}
\caption{Minimum non-mitigated genuine witness values measured for $n$ qubit GHZ states prepared on \belem{}, \toronto{}, \guadalupe{}, and \jakarta{}.}
\label{tab:GHZ_genuine_only}
\end{table}

\begin{table}[h!]
\centering
\begin{tabular}{ |p{2.8cm}|c|c|c|c| } 
\hline
 \multicolumn{5}{|c|}{Biseparable Witness-GHZ orbit} \\ 
\hline
 Name & Qubits & Star & Complete & Method\\ 
 \hline
 \multirow{6}{*}{\texttt{ibmq\_belem}} 
 & 2 & -0.740234 & -0.740234 & \naive{} \\ 
  \cline{2-5}
 & 2 & -0.728271 & -0.728271 & \unitary{} \\
  \cline{2-5}
 \cline{2-5}
 & 3 & -0.644531 & -0.524170 & \naive{} \\ 
  \cline{2-5}
 & 3 & -0.705566 & -0.628662 & \unitary{} \\
  \cline{2-5}
 \cline{2-5}
 & 4 & -0.583984 & -0.225342 & \naive{} \\ 
  \cline{2-5}
 & 4 & -0.540283 & -0.531494 & \unitary{} \\
 \hline
 \multirow{6}{*}{\texttt{ibmq\_toronto}} 
 & 2 & -0.718018 & -0.718018 & \naive{} \\ 
  \cline{2-5}
 & 2 & -0.710693 & -0.710693 & \unitary{} \\
  \cline{2-5}
 \cline{2-5}
 & 3 & -0.668213 & -0.511719 & \naive{} \\ 
  \cline{2-5}
 & 3 & -0.667725 & -0.663574 & \unitary{} \\
  \cline{2-5}
 \cline{2-5}
 & 4 & -0.411133 & -0.002686 & \naive{} \\ 
  \cline{2-5}
 & 4 & -0.408691 & -0.400391 & \unitary{} \\
\hline
 \multirow{6}{*}{\texttt{ibmq\_guadalupe}}
 & 2 & -0.860352 & -0.860352 & \naive{} \\ 
  \cline{2-5}
 & 2 & -0.856201 & -0.856201 & \unitary{} \\
  \cline{2-5}
 \cline{2-5}
 & 3 & -0.716309 & -0.534180 & \naive{} \\ 
  \cline{2-5}
 & 3 & -0.729248 & -0.708740 & \unitary{} \\
  \cline{2-5}
 \cline{2-5}
 & 4 & -0.580322 & -0.275391 & \naive{} \\ 
  \cline{2-5}
 & 4 & -0.496826 & -0.443604 & \unitary{} \\
 \hline
 \multirow{6}{*}{\texttt{ibmq\_jakarta}} 
 & 2 & -0.792969 & -0.792969 & \naive{} \\ 
  \cline{2-5}
 & 2 & -0.779785 & -0.779785 & \unitary{} \\
  \cline{2-5}
 \cline{2-5}
 & 3 & -0.754395 & -0.599365 & \naive{} \\ 
  \cline{2-5}
 & 3 & -0.699707 & -0.687988 & \unitary{} \\
  \cline{2-5}
 \cline{2-5}
 & 4 & -0.710693 & -0.425781 & \naive{} \\ 
  \cline{2-5}
 & 4 & -0.696289 & -0.652100 & \unitary{} \\
 \hline
\end{tabular}
\caption{Minimum non-mitigated bisepable witness values measured for $n$ qubit GHZ states prepared on \belem{}, \toronto{}, \guadalupe{}, and \jakarta{} via the \unitary{} method.}
\label{tab:GHZ_bisep_only}
\end{table}

\section{Pseudocode}
\label{appendix:pseduocode}
We provide pseudocode to implement data collection using the \naive{} method (in Algorithm \ref{alg:naive}) and the \unitary{} method (in Algorithm \ref{alg:unitary}) described in the main text.
\begin{algorithm}[H]
\caption{Circuit construction and data collection for the \naive{} method.}\label{alg:naive}
\begin{algorithmic}
\Require $|\lbrace q_i\rbrace| \geq 2$ \Comment{fixed set of hardware qubits}
\Ensure $\mathcal{B}[\lbrace q_i\rbrace]$ is connected
\State{$\mathcal{G} \gets \mathcal{B}[\lbrace q_i\rbrace]$}
\State{$\lbrace \ell \rbrace \gets []$}
\For{$i \gets 0$ to $2^{n+1}$} \Comment{Generate random LC sequences}
    \State {$m \gets x \sim $Unif$(0,2n)$}
    \State{$\vec{\ell}_i \gets []$}
    \For{$j \gets 0$ to $m$}
    \State{$\vec{\ell}_i$.append($y \sim $Unif$(0,n-1)$)}
    \EndFor
    \State {$\ell$.append($\vec{\ell}_i$)}
    \EndFor
\State {M $\gets$ []}
\For{$i \gets 0$ to $m$}
\State {$\mathcal{G}^{\prime} \gets \mathcal{G}$}
    \For{$j \gets 0$ to $|\vec{\ell}_i|$}
        \State {$\mathcal{G}^{\prime} \gets \ell_i[j](\mathcal{G}^{\prime})$} \Comment{Construct LC transformed graph}
    \EndFor
    \State {$| \psi_{\mathcal{G}^{\prime}}  \rangle \gets \mathcal{U}(\mathcal{G})|0\rangle^{\otimes n}$} \Comment{Prepare state}
    \State {$ \lbrace g^{\prime}_k \rbrace \gets  g_k^{(\scriptG{G}^{\prime})}$} \Comment{define generators (cf. Eq. \ref{eq:generator_def})}
    \For{$j \gets 0$ to $k$} 
            \State{ M.append$(\langle \psi_{\mathcal{G}^{\prime}} | g^{\prime}_j |\psi_{\mathcal{G}^{\prime}}  \rangle$)}
        \EndFor
    \EndFor
\end{algorithmic}
\end{algorithm}
\begin{algorithm}[H]
\caption{Circuit construction and data collection for the \unitary{} method.}\label{alg:unitary}
\begin{algorithmic}
\Require $|\lbrace q_i\rbrace| \geq 2$ \Comment{fixed set of hardware qubits}
\Ensure $\mathcal{B}[\lbrace q_i\rbrace]$ is connected
\State{$\mathcal{G} \gets \mathcal{B}[\lbrace q_i\rbrace]$}
\State{$\lbrace \ell \rbrace \gets []$}
\For{$i \gets 0$ to $2^{n+1}$} \Comment{Generate random LC sequences}
    \State {$m \gets x \sim $Unif$(0,2n)$}
    \State{$\vec{\ell}_i \gets []$}
    \For{$j \gets 0$ to $m$}
    \State{$\vec{\ell}_i$.append($y \sim $Unif$(0,n-1)$)}
    \EndFor
    \State {$\ell$.append($\vec{\ell}_i$)}
    \EndFor
\State {M $\gets$ []}
\For{$i \gets 0$ to $m$}
\State {$| \psi_{\mathcal{G}}  \rangle \gets \mathcal{U}(\mathcal{G})|0\rangle^{\otimes n}$} \Comment{Prepare state}
\State {$\mathcal{U}_{S} \gets \mathbbm{1}$} 
    \For{$j \gets 0$ to $|\vec{\ell}_i|$}
        \State {$\mathcal{U}_{S}$.append($\mathcal{U}(\ell_i[j]))$} \Comment{Construct local unitary (cf. Eq. \ref{eq:generator_transform})}
        \State {$ \lbrace g_k \rbrace \gets \lbrace g^{\prime}_k \rbrace$} \Comment{Transform generators (see Ref. \cite{cabello2009compact})}
    \EndFor                  
    \State{ M.append$(\langle \psi_{\mathcal{G}^{\prime}} |\mathcal{U}_{S}^{\dagger} g^{\prime}_j \mathcal{U}_{S}|\psi_{\mathcal{G}^{\prime}}  \rangle$)}
\EndFor
\end{algorithmic}
\end{algorithm}
\section{Hardware calibration data}
\label{appendix:calibration_data}
\begin{figure}[htbp]
  \centering
  \includegraphics[width=\columnwidth]{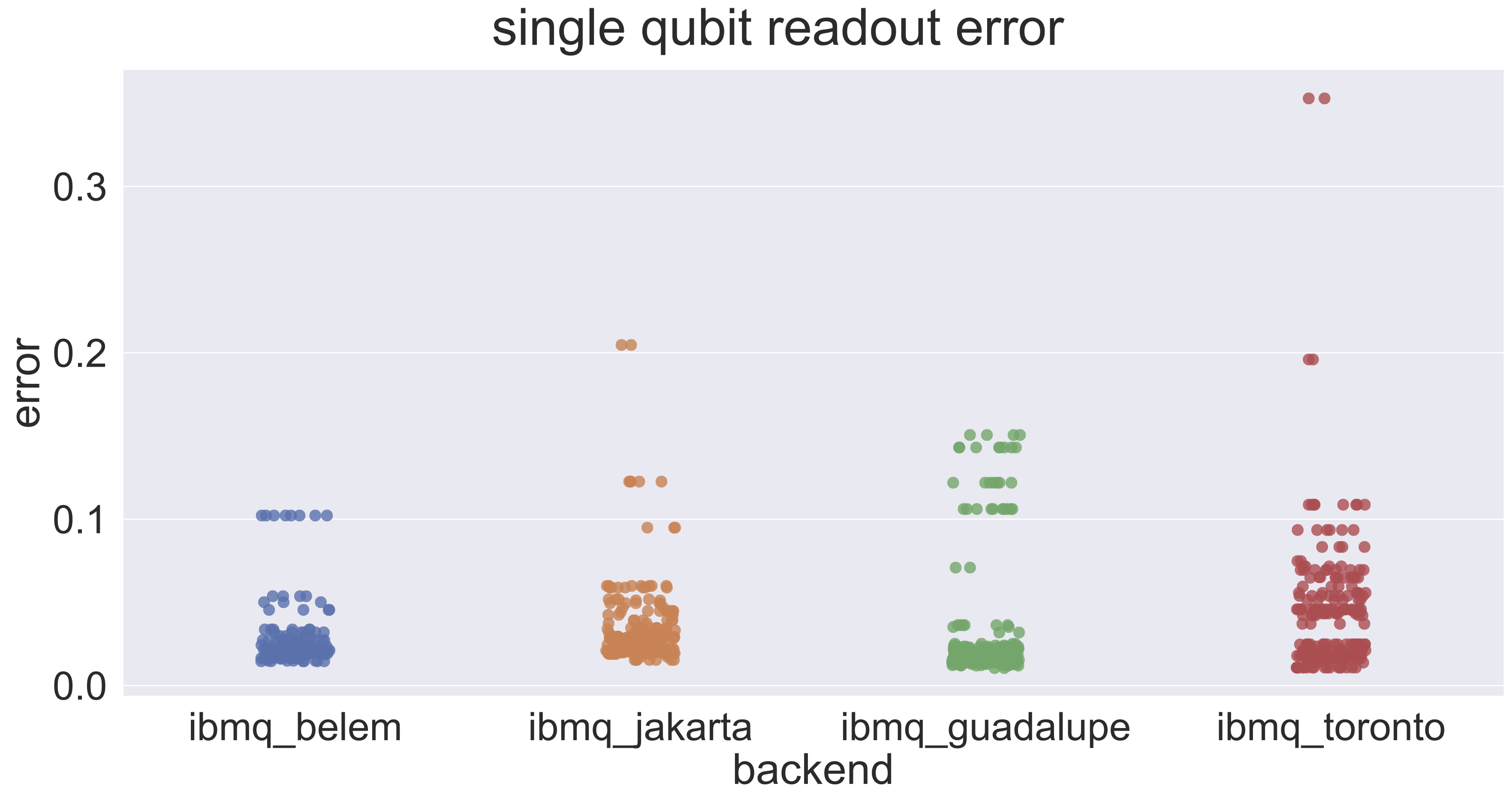}
  \caption{Single qubit readout error values collected from calibration data for each backend.  Only the error values for the individual hardware qubits used in our data collection are plotted.  Calibration data was pulled over several days.}
  \label{fig:readout_error_calibration_data}
\end{figure}

\begin{figure}[htbp]
  \centering
  \includegraphics[width=\columnwidth]{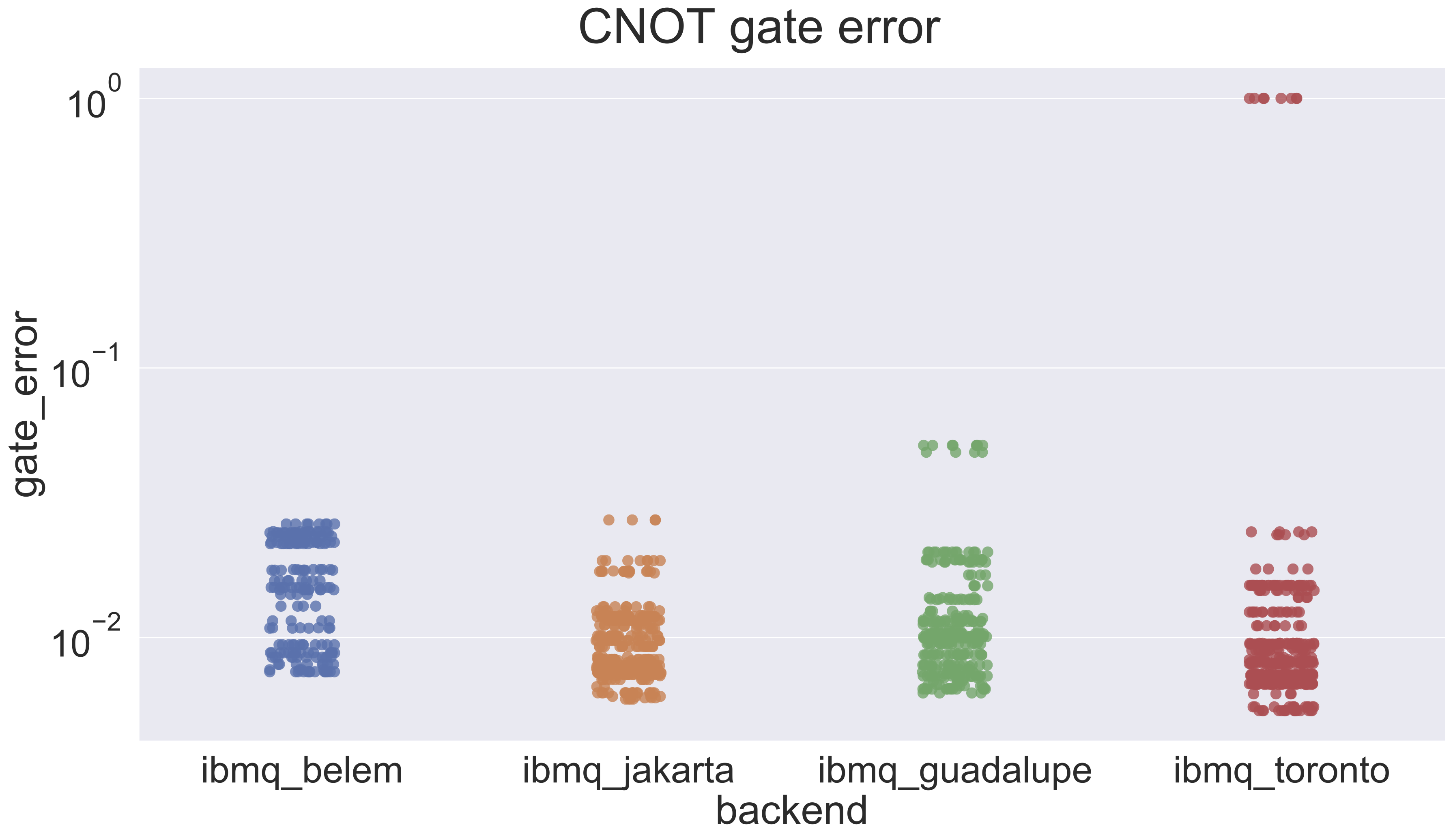}
  \caption{CNOT gate error values collected from calibration data for each backend plotted on log scale.  Only the error values for gates on the individual hardware qubits used in our data collection are plotted.  Calibration data was pulled over several days.}
  \label{fig:cx_gate_error_calibration_data}
\end{figure}

Once the circuit batches are composed (see Section \ref{sec:bulk_data}), they are sent to cloud-based queues.  Prior to sending the job batches, we pull the most recent calibration data available for our target backend, and store this data.  This data includes single qubit calibration data ($T_1, T_2$ values, anharmonicity, frequency) gate calibration data (length, frequency) and also error characterization (single qubit readout error and individual gate error).  Based on the features we analyzed in the main text, the efficacy of readout error mitigation and the correlation with number of CNOT gates, we present in Figs. \ref{fig:readout_error_calibration_data},\ref{fig:cx_gate_error_calibration_data} the values of the single qubit readout error and CNOT gate error reported by IBM for each backend, collected at the time of data collection.

In Fig. \ref{fig:stabilizer_quartiles} of the main text, the two-qubit stabilizer measurements on \toronto{} had a large negative skew. We highlight the connection between the qubit subsets that resulted in low stabilizer measurements and the low-level calibration data extracted here. As a threshold we isolate the stabilizer measurements that are below the noise robust threshold of the genuine entanglement witness ($1/2$). This data was collected in April of 2022.

For this subset of data, the couplers which correspond to low stabilizer measurements were $[0, 1], [2, 3] ,[3, 5]$. The recorded CNOT gate errors for couplers $[0,1], [1,0], [2,3],[3,2]$ was $1.0$, but for couplers $[3,5], [5,3]$ the gate error was reported as $0.024$.  For the individual qubits, the reported readout errors were: $0.353$ for qubit 0, $0.04$ for qubit 1,  $0.02$ for qubit 2, $0.04$ for qubit 3, and $0.02$ for qubit 5. While the high gate error associated with couplers $[0,1], [1,0], [2,3],[3,2]$, and high readout error for qubit 0 may have contributed to the poor preparation of a two-qubit entangled state (Bell state) on hardware qubits 0 and 1, the calibration data for qubits $3$, $5$ and the associated couplers $[3,5], [5,3]$ does not correlate with the observed low stabilizer measurements.

\section{Minimum Entanglement Witness Values}
\label{appendix:min_witness}
Figures \ref{fig:bisep_entanglement_heatmaps_jakarta_toronto}, \ref{fig:bisep_entanglement_heatmaps_guadalupe_belem} in the main text show the median witness value evaluated over several graph states drawn from the relative LC graph orbit of $n$ qubit subgraphs.  In Table \ref{tab:best_gen_witness} we report the lowest non-mitigated genuine entanglement witness values measured for $n$ qubits on each backend. Then, in Table \ref{tab:best_bisep_witness} we report the lowest biseparable witness values evaluated on each backend, using the \naive{} or the \unitary{} method.  When reporting the hardware qubit subset, we emphasize the pair of qubits used to define the generators used in the specific witness.

\begin{table*}[h!]
\centering
\begin{tabular}{ |p{2.3cm}|p{4.5cm}| c | p{4.5cm} |c | } 
\hline
\multicolumn{5}{|c|}{Genuine Witness} \\ 
\hline
 Name & Qubits & \unitary{} & Qubits & \naive{} \\ 
 \hline
 \multirow{4}{*}{\texttt{ibmq\_belem}}
 & [0,1] & -0.72827 & [0,1] & -0.740234 \\ 
 \cline{2-5}
&  [0, 1, 2] & -0.541260 & [0, 1, 2] & -0.458252\\ 
 \cline{2-5}
&  [0, 1,2, 3]  & -0.135742 & [0, 1,2, 3]  & -0.127197\\ 
 \cline{2-5}
& [0, 1, 2, 3, 4] & 0.371826 & [0, 1, 2, 3, 4] & 0.333252\\ 
 \hline
 \multirow{8}{*}{\texttt{ibmq\_toronto}} 
 & [1, 2] & -0.710693 & [1, 2] &-0.718017 \\ 
 \cline{2-5}
& [5, 8, 11] &  -0.471680 & [5, 8, 11] & -0.476318 \\ 
 \cline{2-5}
& [5, 8, 11, 14] & -0.143555 & [5, 8, 11, 14] & -0.186523\\ 
 \cline{2-5}
& [3, 5, 8,11, 14] & 0.241211 & [3, 5, 8,11, 14] & 0.275879 \\ 
 \cline{2-5}
& [3, 5, 8,11, 14, 16] & 0.659424 & [3, 5, 8,11, 14, 16] & 0.584717\\ 
 \cline{2-5}
& [3, 5, 8, 11, 14, 16, 19] & 1.558837 & [3, 5, 8, 11, 14, 16, 19] & 1.526367\\ 
 \cline{2-5}
& [0, 1, 2, 4, 7, 6, 10, 12] & 2.520996 & [3, 5, 8, 11,14, 16, 19, 22] & 2.414062\\ 
 \cline{2-5}
& [3, 5, 8, 11, 14, 16, 19, 22, 25] & 3.303223 & [3, 5, 8, 11, 14, 16, 19, 22, 25] & 3.167969\\ 
\hline
 \multirow{6}{*}{\texttt{ibmq\_guadalupe}} 
 & [3, 5] & -0.85620 & [3, 5] & -0.86035\\ 
 \cline{2-5}
&   [0, 1, 2] & -0.587158 & [0, 1, 2] & -0.568359\\ 
 \cline{2-5}
&[10, 12, 6, 7]& -0.495117 & [10, 12, 15, 13] & -0.513427\\ 
 \cline{2-5}
& [7, 10, 12, 13, 14] & 0.070801 & [7, 10, 12,13, 14] & 0.099609\\ 
 \cline{2-5}
& [5, 8, 11, 12, 13, 14] & 0.267822 & [0, 1, 2,4, 6, 7] & 0.102051\\ 
 \hline
 \multirow{6}{*}{\texttt{ibmq\_jakarta}} 
 & [1, 3] & -0.779785 & [1, 3] &-0.792969\\ 
 \cline{2-5}
& [0, 1, 2]  &  -0.552246 & [0, 1, 2] & -0.640381\\ 
 \cline{2-5}
& [0, 1, 3, 5] &   -0.429688 & [0, 1, 2, 3] &-0.447998\\  
 \cline{2-5}
& [0, 1, 2, 3, 5] &   -0.028564 & [0, 1, 2,3, 5] & -0.198242\\ 
 \cline{2-5}
& [0, 1, 2, 3, 5, 6] & 0.47729 &[0, 1, 2, 3, 5, 6] & 0.186279\\ 
 \cline{2-5}
& [0, 1, 2, 3,5, 4, 6] & 1.264648 & [0, 1, 2, 3, 5, 4, 6] & 1.053223\\ 
 \hline
\end{tabular}
\caption{Hardware qubits that returned the minimum non-mitigated genuine witness values measured on \texttt{ibmq\_belem}, \texttt{ibmq\_toronto}, \texttt{ibmq\_guadalupe}, and \texttt{ibmq\_jakarta}.}
\label{tab:best_gen_witness}
\end{table*}

\begin{table*}[h!]
\centering
\begin{tabular}{ |p{2.3cm}|p{4.5cm}| c | p{4.5cm} |c | } 
\hline
 \multicolumn{5}{|c|}{Biseparable Witness} \\ 
\hline
 Name & Qubits & \unitary{} & Qubits & \naive{} \\ 
 \hline
 \multirow{4}{*}{\texttt{ibmq\_belem}} 
 & [\textbf{0,1}] & -0.72827 & [\textbf{0,1]} & -0.74023 \\ 
 \cline{2-5}
&[0, \textbf{1,2}] & -0.7056 & [0, \textbf{1,2}] &-0.6445\\ 
 \cline{2-5}
& [2, 1, \textbf{3,4}] & -0.5991 & [2, 1, \textbf{3,4}] & -0.6057\\ 
 \cline{2-5}
& [0, 1, 2, \textbf{3,4}] & -0.5630 & [0, 1, 2, \textbf{3,4}] & -0.6086 \\ 
 \hline
 \multirow{8}{*}{\texttt{ibmq\_toronto}} 
 & [\textbf{1,2}] &-0.710693 & [\textbf{1,2}] & -0.718017 \\ 
 \cline{2-5}
& [5, \textbf{8,11}] & -0.667725 & [5, \textbf{8,11}] &-0.668213 \\ 
 \cline{2-5}
& [5, 8, \textbf{11,14}] & -0.709717 & [5, 8, \textbf{11,14}] & -0.697510 \\ 
 \cline{2-5}
& [3, 5, 8, \textbf{11,14}] & -0.737793 & [3, 5, 8, \textbf{11,14}] &-0.737549 \\ 
 \cline{2-5}
& [3, 5, 8, 11, \textbf{14,16}] & -0.747559 & [3, 5, 8, 11, \textbf{14,16}] &-0.754639\\ 
 \cline{2-5}
& [3, 5, 8, \textbf{11,14}, 16, 19] & -0.600342 & [3, 5, 8, 11, 14, \textbf{16,19}] & -0.632324\\ 
 \cline{2-5}
& [3, 5, 8, 11, \textbf{14,16}, 19, 22] & -0.501953 & [3, 5, 8, \textbf{11,14}, 16, 19, 22] &-0.506348\\ 
 \cline{2-5}
& [3, 5, 8, 11, 14, 16, 19, \textbf{22,25}] & -0.645264 & [3, 5, 8, 11, 14, 16, 19, \textbf{22,25}] &-0.635742\\ 
\hline
 \multirow{6}{*}{\texttt{ibmq\_guadalupe}} 
 & [\textbf{3,5}] & -0.856201 &[\textbf{3,5}] &-0.860352\\ 
 \cline{2-5}
& [\textbf{0,1}, 2] & -0.729248 & [\textbf{0,1}, 2] &-0.716309\\ 
 \cline{2-5}
&[11, \textbf{12,13}, 14] & -0.771729 & [11, \textbf{12,13}, 14] &-0.786621\\ 
 \cline{2-5}
& [7, 10, 12, \textbf{13,14}] & -0.730957 & [7, 10, 12, \textbf{13,14}] & -0.736816\\ 
 \cline{2-5}
& [5, 8, 11, \textbf{12,13}, 14] & -0.779297 & [0, 1, 2, 4, \textbf{6,7}] &-0.793213\\ 
 \hline
 \multirow{6}{*}{\texttt{ibmq\_jakarta}} 
 & [\textbf{1,3}] & -0.779785 & [\textbf{1,3}] &-0.792969\\ 
 \cline{2-5}
& [0, \textbf{1,2}] & -0.699707 & [0, \textbf{1,2}] & -0.754395\\ 
 \cline{2-5}
& [0, 1, \textbf{3,5}] &  -0.759277 & [0, 1, \textbf{3,5}] & -0.762695\\  
 \cline{2-5}
& [0, 1, 2, \textbf{3,5}] & -0.719971 & [0, 1, 2, \textbf{3,5}] &-0.753174\\ 
 \cline{2-5}
& [0, 1, 2, 3, \textbf{5,6}] & -0.707031 & [0, 1, 2, 3, \textbf{5,6}] & -0.749023\\ 
 \cline{2-5}
& [0, 1, 2, 3, \textbf{5,4}, 6] & -0.597900 & [0, 1, 2, 3, \textbf{5,4}, 6] &-0.5551758\\ 
 \hline
\end{tabular}
\caption{Minimum non-mitigated biseparable witness values measured for $n$ qubits on \texttt{ibmq\_belem}, \texttt{ibmq\_toronto}, \texttt{ibmq\_guadalupe}, and \texttt{ibmq\_jakarta}.}
\label{tab:best_bisep_witness}
\end{table*}

\section{Correlation Matrices}
\label{appendix:pearson_correlation}
\begin{figure*}[htbp]
    \centering
    \includegraphics[angle=0,width=\textwidth]{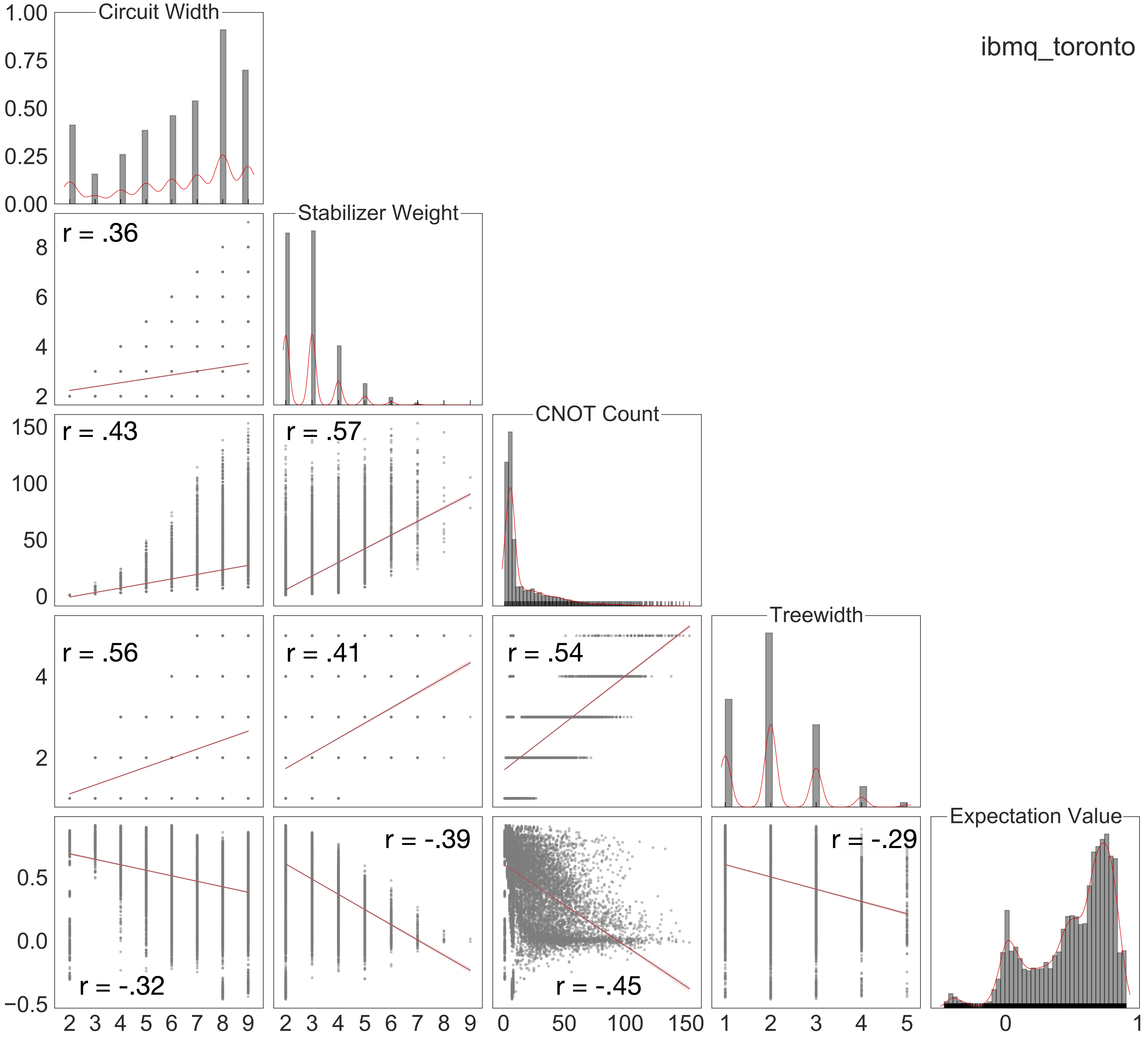}
    \caption{Correlation matrix for \toronto{}.  Linear regression fit plotted in red with confidence interval and Pearson's r-coefficient computed between features are reported on individual scatter plots. \toronto{} r-coefficients are computed with 19272 degrees of freedom.}
  \label{fig:toronto_correlation_matrix}
\end{figure*}

\begin{figure*}[htbp]
    \centering
    \includegraphics[angle=90,width=\textwidth]{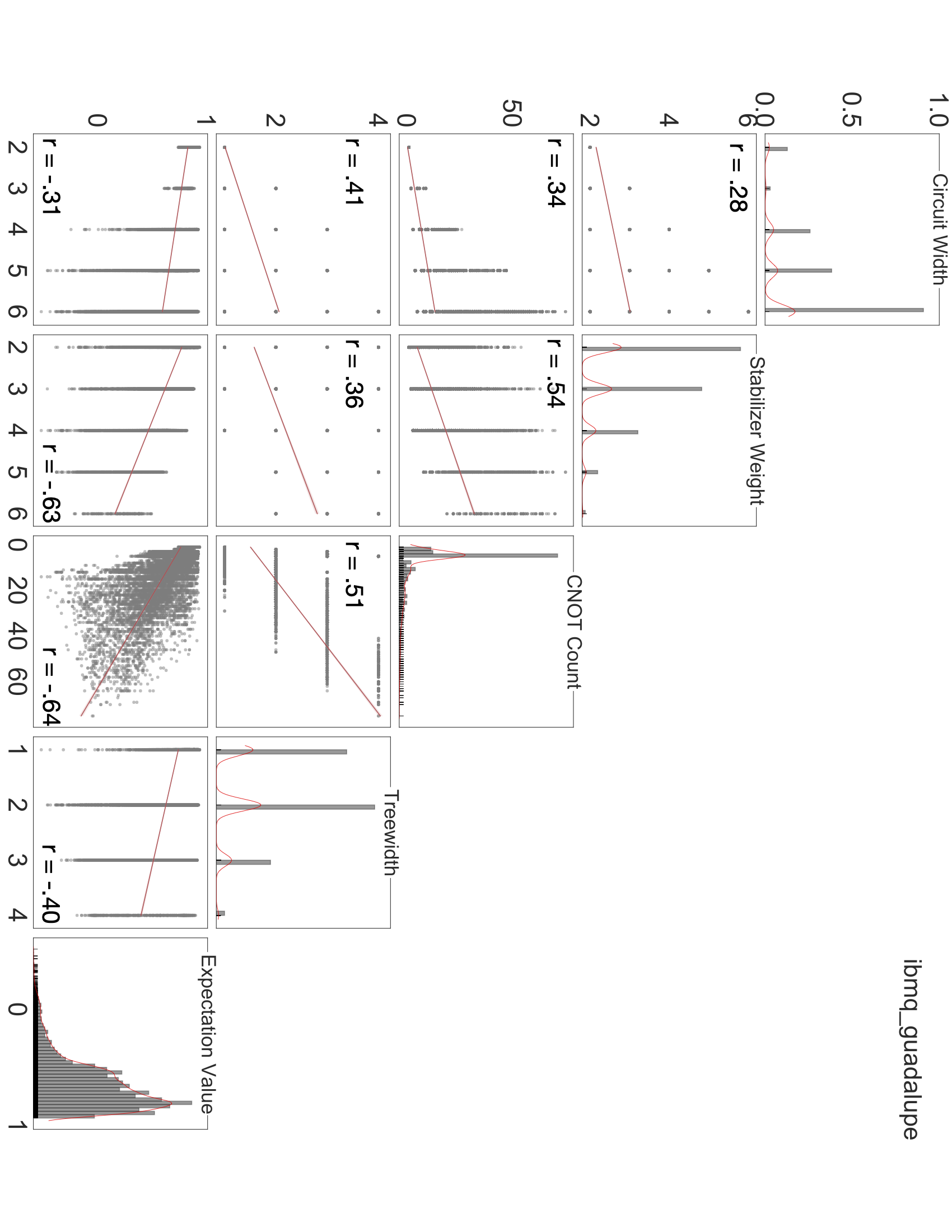}
    \caption{Correlation matrix for \guadalupe.  Linear regression fit plotted in red with confidence interval and Pearson's r-coefficient computed between features are reported on individual scatter plots. \guadalupe{} r-coefficients are computed with 17368 degrees of freedom.}
  \label{fig:guadalupe_correlation_matrix}
\end{figure*}

\begin{figure*}[htbp]
    \centering
    \includegraphics[angle=90,width=\textwidth]{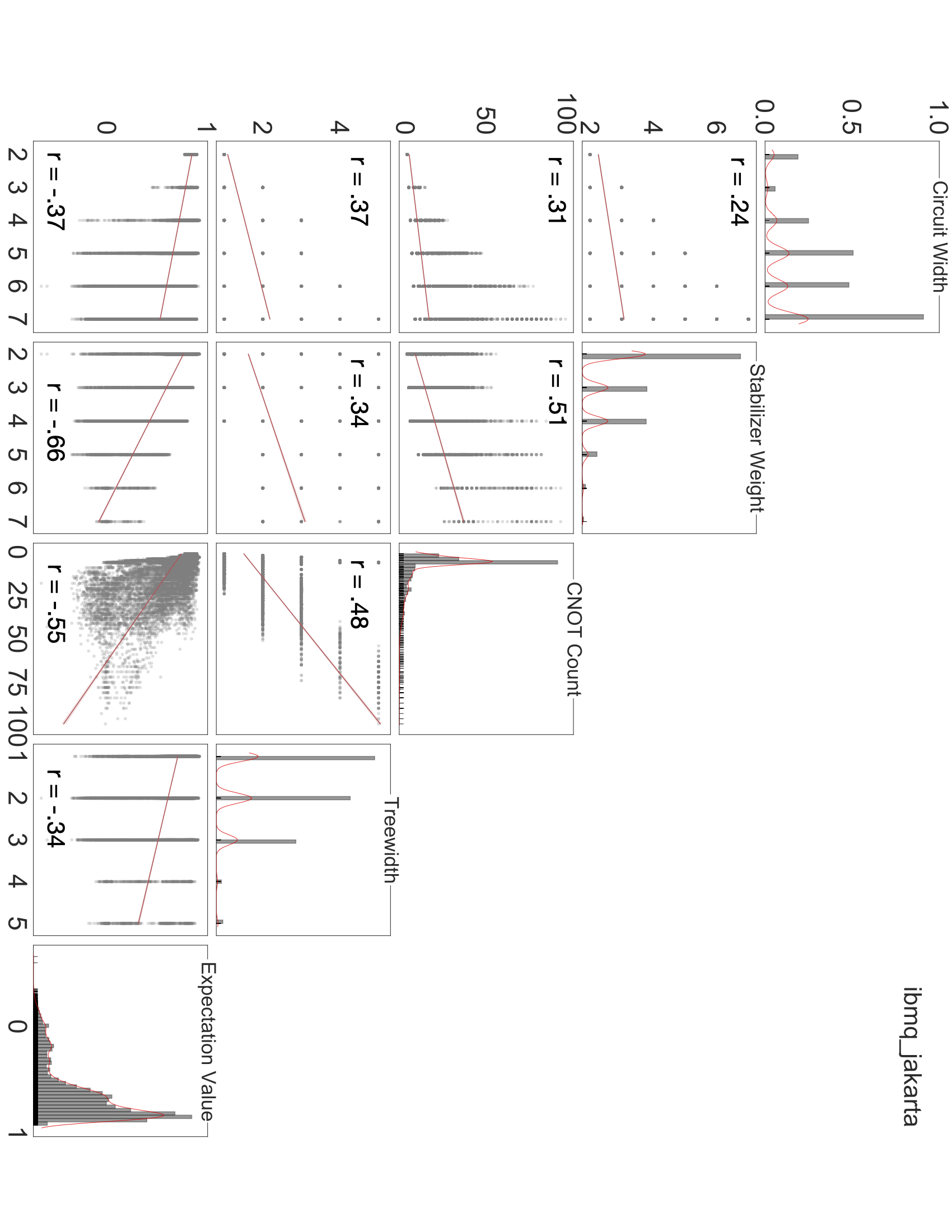}
    \caption{ Correlation matrix for \jakarta{}. Linear regression fit plotted in red with confidence interval and Pearson's r-coefficient computed between features are reported on individual scatter plots. \jakarta{}  r-coefficient values were computed with 23018 degrees of freedom.}
  \label{fig:jakarta_correlation_matrix}
\end{figure*}

\begin{figure*}[htbp]
    \centering
    \includegraphics[angle=90,width=\textwidth]{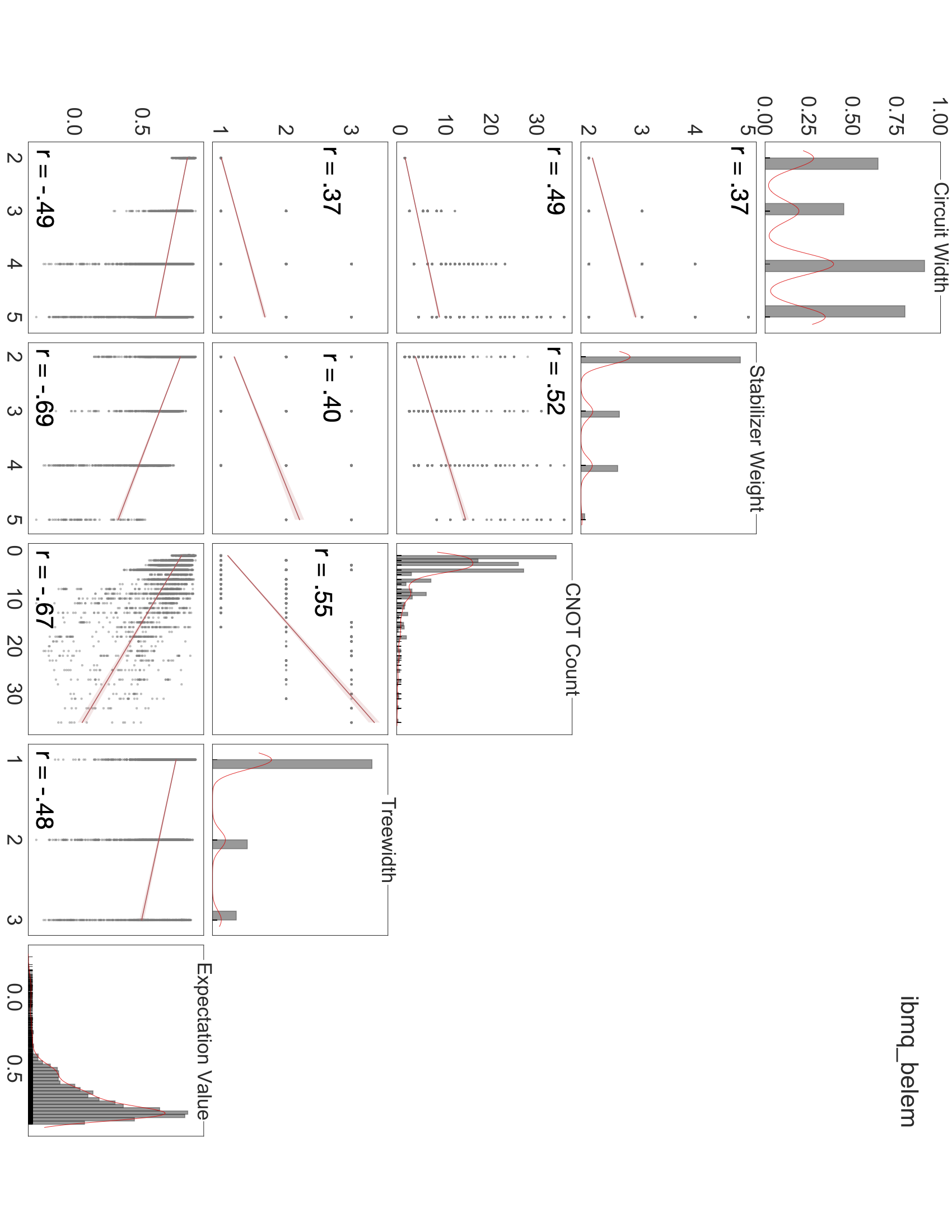}
    \caption{Correlation Matrix for \belem. Linear regression fit plotted in red with confidence interval and Pearson's r-coefficient computed between features are reported on individual scatter plots. \belem{} r-coefficient values were computed with 5650 degrees of freedom.}
  \label{fig:belem_correlation_matrix}
\end{figure*}

In the main text Section \ref{sec:discussion} we structure our analysis along several circuit structure features: circuit width, number of  in the transpiled circuits, graph treewidth, and stabilizer weight.  These features were chosen based on the observed correlations with stabilizer expectation values. For each feature, we compute the Pearson r-coefficient, and show correlation matrix plots for each backend in Figs.~\ref{fig:toronto_correlation_matrix},\ref{fig:guadalupe_correlation_matrix},\ref{fig:jakarta_correlation_matrix},\ref{fig:belem_correlation_matrix}. All r-coefficient values were found with with $p<0.001$.

\end{document}